\documentclass[runningheads]{llncs}
\usepackage[T1]{fontenc}
\usepackage[numbers]{natbib}

\usepackage{amsmath}
\usepackage{url,hyperref,cleveref}
\usepackage{booktabs}
\usepackage{setspace}
\usepackage{graphicx}
\usepackage{balance}
\usepackage{subcaption}

\begin{document}
\title{Emergent Dynamics in Heterogeneous Life-Like Cellular Automata}
%
%


\author{Aarati Shrestha\inst{1} \and
Felix Reimers\inst{1} \and
Sanyam Jain\inst{1} \and
Paolo Baldini\inst{2} \and
Michele Braccini\inst{2} \and
Andrea Roli\inst{2} \and
Stefano Nichele\inst{1,3}
}
\authorrunning{A. Shrestha et al.}
%
\institute{Østfold University College, Halden, Norway \and
University of Bologna, Cesena, Italy \and
Oslo Metropolitan University, Oslo, Norway\\
\email{\{felix.s.reimers, stefano.nichele\}@hiof.no}}
\maketitle              
\begin{abstract}
The Game of Life (GoL), one well known 2D cellular automaton, does not typically ensure \textit{interesting} long-term phenotypic dynamics. Therefore, while being Turing complete, GoL cannot be said to be open-ended. In this work, we extend GoL with the opportunity for local mutations, thus enabling a heterogeneous life-like cellular automaton guided by an evolutionary inner loop. Additionally, we introduce the concept of cell ageing to ensure that cell aliveness (activated by inheritance with variation, and controlled by ageing) and actual cell computation (governed by life-like rules on local neighborhoods) are kept conceptually separated. We conduct an experimental campaign to identify suitable parameters that produce long-term phenotypic dynamics and favor genotypic innovations.  

\end{abstract}
\section{Introduction}

Conway's Game of Life (GoL) \citep{gardner1970mathematical, berlekamp2004winning} is a very well known cellular automaton (CA) that has been proven to be Turing complete \citep{rendell2002turing}. Therefore, in principle, it can execute a universal constructor or any Turing machine. GoL has been widely used as a model system to study the emergence of complex behaviors from simple rules \citep{bak1989self, suzuki2008homeodynamics}. In fact, the exploration of complex moving structures, such as spaceships (complex gliders) that can carry information and encode computations, is an active area of research \citep{eppstein2000searching}. Additionally, gliders have been studied in relation to agency \citep{biehlgame}. However, such computational structures are rather fragile in that even small perturbations will destroy the precisely handcrafted computations. Additionally, from a random initial configuration typically GoL settles into a rather boring and not particularly useful behavior, as most of the living cells die out besides a few persistent or oscillating structures. Therefore, while in principle GoL supports arbitrarily complex computations, in practice it does not appear to be open-ended. Open-endedness is a property of a system, observed for example in biological systems, that ensures never ending innovation and discovery of novel solutions, thus providing continuously increasing complexity \citep{packard2019overview}. 

In order to allow more interesting behaviors to emerge, McCaskill and Packard \citep{mccaskill2019analysing,packard2024open} investigated a GoL world where local mutations diversify GoL rules to other life-like rules, named genelife. Genelife created more diversity in the local dynamics and therefore allowed the emergence of different kinds of non-uniform gliders and structures. Other attempt at coupling life-like rules and evolution are Sprout Life \citep{sproutlife} and evolife \citep{evolife}. On the other hand, in such CA worlds there is no distinction between the aliveness of cells (agents) and their cellular state. In fact, aliveness and computation carry the same meaning. In other words, there is no distinction between the agent and the computation it carries, i.e. aliveness is the computation. However, it may be beneficial to differentiate the conditions for cells to become alive and survive (and therefore generate offspring through inheritance and variation) while keeping the actual computation based on local information conceptually separated. 

Medernach et al. introduced the concept of cell age in their HetCA system \citep{medernach2013long, medernach2016evolution}, while using evolved genetic programming (GP) rules as mechanisms to update the states of cells based on local neighboroods. In their work, each cell possesses an age counter that increments at each time step. Cells can live only for a certain number of steps and during those steps they can update their states through their local rules. When the age limit is reached, cells transition to the decay state, i.e., a living state where cells do not update their state any longer (however their state is still available to their neighbors). After the decay phase, cells die and empty their locations. Empty locations that are neighbors of living cells may probabilistically become alive by inheriting the GP rules from a living neighbor through mutation. As such, HetCA produces interesting long term phenotypic dynamics without any sign of stagnation or repetitive behavior. 

One open question is whether a more simplified rule-set, such as life-like rules (instead of intricate GP programs), can support similar long-term phenotypic and genotypic dynamics by incorporating the concept of ageing, together with an evolutionary inner loop. 
Additionally, the concept of age allows for incorporating environmental factors such as "energy", which may increase/decrease the lifespan of cells. This may be favourable for simulating different ecologies, as well as plugging-in tasks to be solved in the form of "rewards and penalties". 
One may imagine a substrate that initially supports long-term evolutionary dynamics (without a goal). Then energy is introduced and this life-like "solver" adapts to the environmental conditions. Finally, the task is removed and open-ended life continues. In this work, we experiment to find out the ideal conditions to support long-term dynamics in heterogeneous life-like CA with age constrains and local evolution. 

\section{Background}

\subsection{Game of Life Cellular Automaton}

\begin{figure}[t]
    \centering
    \includegraphics[width=0.55\linewidth]{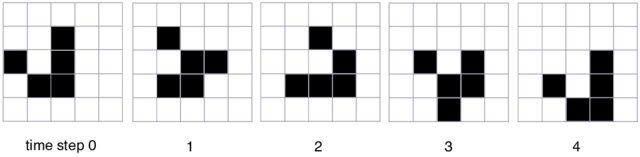}  
    \caption{States of a glider in Conway's Game of Life}
    \label{fig:glider_gol}
\vspace{-10pt}
\end{figure}

Conway's Game of Life (GoL) is a 2D Cellular Automaton where each cell in the grid is either in an alive or dead state. The state of a cell is determined by its eight surrounding neighbours, known as the Moore Neighbourhood, and follows a simple transition rule denoted as B3/S23. This notation signifies that a cell is born if it has precisely 3 live neighbours, and it survives if it has 2 or 3 live neighbours; otherwise, it dies. Figure \ref{fig:glider_gol} depicts the subsequent stages of the GoL's glider pattern, showcasing the intricate interplay of simple rules leading to emerging dynamics.

While extensive research has been conducted into the exploration of 2D GoL, it is crucial to acknowledge the parallel efforts dedicated to investigating higher dimensions. Notably, Bays \citep{bays1987candidates} contributed significantly by extending GoL in three dimensions, introducing variations in the transition rules.

\subsection{Life-Like Cellular Automata}

Life-like CA (signifying that they are similar to the Game of Life) allow for variations of the GoL state transition function. In particular, for all CA in this family, the new state of a cell can be represented by a function of the number of neighbors that are currently alive (including the cell's own state). As for GoL, a typical notation of a life-like rule would consist in a string denoted by B (born) followed by the number of neighbors (from 0 to 9), and then a string denoted by S (survive), also followed by the number of neighbors (from 0 to 9) \citep{wuensche2011exploring}. Therefore, there exist a total of $2^{18} = 262,144$ rules \citep{eppstein2010growth}. Only a few have been extensively explored \footnote{\url{https://conwaylife.com/wiki/List_of_Life-like_rules}}. For a recent exploration of complexity in life-like rules, please refer to \citep{pena2021life}.

\subsection{Heterogeneous and Evolving CA}

While typically the state transition function governing state updates of all cells in the CA is homogeneous, i.e., all cells are governed by the same rules, in heterogeneous CA each cell has its own (potentially unique) rule. Such rules may also vary over time. As such, the space of possible CA becomes rather vast and artificial evolution is often used to search for suitable rules \citep{sipper1998computing, sipper1996co, sipper1999computation, vichniac1986annealed}. 

Evolution may be used as an outer loop, where a genotype consists of the concatenation of the rules for all cells in the CA. In this case, once the CA phenotype (the actual CA execution for a certain number of steps) is evaluated by a fitness function, the genetic operators affect the overall (concatenated) genotype. Subsequently, a new CA execution is carried out and the results are further evaluated. 

In the case of an inner evolutionary loop, however, genomes encoding cell rules evolve locally and in an online fashion. This means that inheritance with variation is governed by local factors and happens from quiescent (dead) cells to alive cells while the CA computation is being executed. Examples of inner-loop CA evolution include \citep{medernach2013long, medernach2016evolution, mccaskill2019analysing, gregor2021self, randazzo2023biomaker}.



\section{Methodology}

Our model, described in detail in this section, consists of a 2-dimensional heterogeneous cellular automaton where each cell is updated according to distinct life-like rules that operate in local neighbourhoods. Additionally, an inner evolutionary loop controls the local evolution of life-like rules by allowing cells to grow into empty nearby sites through inheritance with variation. At the end of this section, we describe how genotypic and phenotypic diversity are measured.

\subsection{CA Substrate}

\begin{figure}[t]
    \centering
    \includegraphics[width=0.6\textwidth]{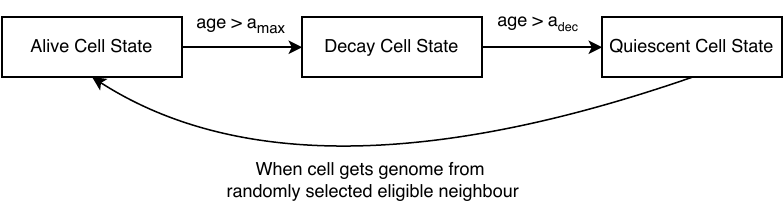}  
    \caption{Life cycle of a cell's \textit{Cell State}.}
    \label{fig:cell_life_cycle}
\vspace{-10pt}
\end{figure}

The CA substrate consists of a 2D grid of cells with a tripartite \textit{Cell State} life cycle, i.e., alive, decay, quiescent (inspired by \cite{medernach2013long}). The CA is initialized with cells either in an alive or quiescent state. The alive cells are those cells that actively apply their \textit{Transition Rule} (genome) to change their Boolean \textit{Grid State} based on a Moore's neighborhood. Quiescent cells are temporarily dead cells awaiting to become alive, therefore they do not have a genome. When initialized, all alive cells have identical Game-of-Life genomes. As the cellular automaton iterates through generations, the cell's \textit{Cell State} undergoes a sequential transformation, transitioning cells from alive states to decay and ultimately to quiescent states. A cell cannot transition directly from decay to alive or from quiescent to decay (nor from alive to quiescent). Instead, these states follow an unidirectional progression, as illustrated in figure \ref{fig:cell_life_cycle}.

The \textit{Cell State} of a given cell is governed by the age of the cell, specifically by two parameters: $a_{max}$ and $a_{dec}$. The $a_{max}$ represents the age threshold for remaining in an alive state. Once a cell's age crosses this threshold, it transitions into a decay state. Cells in a decay state are neither alive nor dead. They hold their genomes but do not apply it to change the \textit{Grid State}. When the age of a decay cell crosses the $a_{dec}$ age threshold, they become quiescent and their age is reset to 0. Quiescent cells may at some point become alive by inheriting a mutated genome from a random alive neighbour. When a cell becomes alive, its age counter starts from 1 and is incremented by one at each CA iteration.

Further, alive cells get their genome when they are born and their genome remains unchanged throughout their life cycle. Cells in decay (and quiescent) state do not apply any transition rule and therefore do not update their \textit{Grid State}. Cells in decay state retain an unaltered \textit{Grid State} (either 0 or 1) throughout the duration of their decay period, while quiescent are considered to be in a \textit{Grid State} 0.

\begin{figure}[t]
    \centering
    \includegraphics[width=0.6\textwidth]{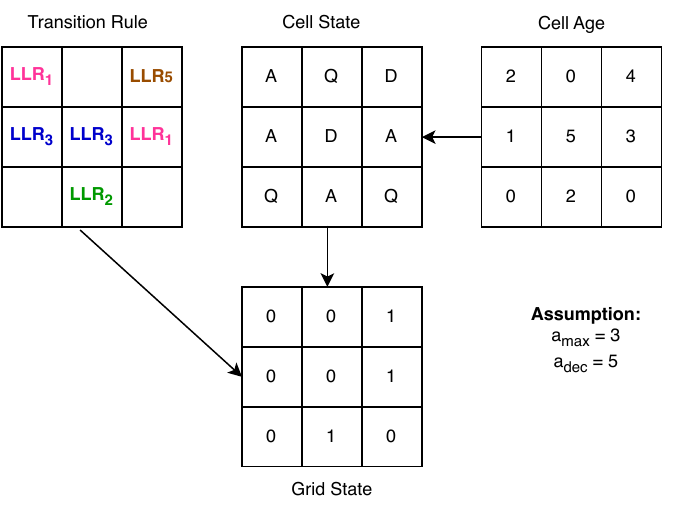}  
    \caption{Example of a 3x3 CA substrate, where the genotype is represented by the \textit{Transition Rules} and the phenotype is composed by three components: \textit{Cell Age} (a counter), \textit{Cell State} (alive, decay, quiescent), and \textit{Grid State} (Boolean cell state representing the ongoing computation).}
    \label{fig:ca_layer}
\vspace{-10pt}
\end{figure}

Figure \ref{fig:ca_layer} illustrates the composition of the CA substrate on a small 3x3 grid. The genomes of each cell are represented by the \textit{Transition Rule} layer, where each cell contains a possibly different Life-Like Rule (LLR). The \textit{Cell State} layer represents whether cells are alive, decaying, or quiescent, while the \textit{Cell Age} layer indicates how many iterations have passed since they became alive. Finally, the \textit{Grid State} layer indicates the actual computation, which consists of a Boolean state that results from the execution of the LLR on the corresponding cell based on the local (Moore) neighborhood.   


\subsection{Evolutionary Inner Loop}
Life-like rules may evolve over time by a mechanisms that allows alive cells to replicate their mutated genome into empty (i.e., quiescent) cells. This process is governed by an inheritance probability parameter that acts on all quiescent cells that have at least a living neighbor. A quiescent cell transitioning to an alive state inherits therefore the genome from a randomly selected alive neighbouring cell. The received genome, which consists of a life-like rule, may undergo mutation. A mutation consist of one of the following genome changes:
\begin{itemize}
  \item A symbol is added, removed, or changed from the B section of the genome;
  \item A symbol is added, removed, or changed from the S section of the genome.
\end{itemize}
Legitimate life-like genomes range from empty $B\cdot S \cdot$ strings to $B0123456789 \ S0123456789$. This means that there are $2^{18} = 262,144$ possible combinations. Appropriate checks are carried out to avoid illegal genomes (for example duplicated symbols).
As an example, a GoL genome represented by the string B2S23 (a cell is born when exactly 2 neighbors are alive, a cell survives when 2 or 3 neighbors are alive, otherwise the cell dies) can be mutated as follows: 
\begin{itemize}
  \item A random symbol is added to the B section: B2\textbf{9}S23
  \item A random symbol is removed from the B section: BS23
  \item A random symbol is changed in the B section: B\textbf{6}S23
  \item A random symbol is added to the S section: B2S23\textbf{9}
  \item A random symbol is removed from the S section: B2S3
  \item A random symbol is changed in the S section: B2S\textbf{1}3
\end{itemize}

It is worth to highlight that we have kept the same life-like notation, i.e., B = born and S = survive. However, in our substrate B and S rules \textbf{do not} governe the birth and survival of cells, since the \textit{Cell State} (alive, decay, quiescent) is only controlled by the \textit{Cell Age} progression. Therefore, B and S in fact control the computational state update represented by the CA \textit{Grid State}. The actual meaning of B is whether a cell in \textit{Grid State} 0 should transition to state 1, while S indicates whether a cell in \textit{Grid State} 1 should remain in state 1.

\subsection{Phenotypic and Genotypic Measures}

We aim at measuring the variation of phenotypes and genotypes over time. The following Tables \ref{tab:qual_measures} and \ref{tab:quant_measures} give an overview of the qualitative and quantitative measures used, respectively, together with the corresponding color palettes.

The genotype of each cell is represented by its life-like rule. A qualitative overview of the current genotypic variability over the entire CA can be visually represented by a 2D-grid, where cells with identical genomes are represented by equal colors. Quiescent cells (a) do not possess any genome and are represented in white, while GoL genomes (b) are always represented in yellow. Other unique genomes (c) are represented by one unique color. 

To quantitatively measure the overall genotypic variation, we count the total (cumulative) number of unique life-like rules discovered during each generation. This gives an indication on whether the CA substrate keeps discovering new genotypes over time. 


The phenotype variation is qualitatively visualized with two 2D grids representing:
\begin{itemize}
    \item the \textit{Cell State} and age - quiescent (a) in white, decay (b) in orange, alive (c) in different green color shades indicating their current \textit{Cell Age} counters,
    \item the actual \textit{Grid States} ($0$, $1$) - quiescent cells (a) are depicted in white, $1$ (b) in green, and $0$ (c) in red.
\end{itemize} 

To quantitatively assess the phenotypic variation over time, we are employing various metrics. 

In order to quantitatively measure fluctuations of the Grid State, the number of cells that change state (either from 0 to 1, or from 1 to 0) between two consecutive iterations are measured. This indicates the phenotypic movement between two consecutive generations. 



Another approach involves tallying the number of living cells in state 1 (living in this case includes both alive and decay cells), the number of alive cells in state 0,  and dead (quiescent) cells, across each generation. This method offers insight into the overall computation in the substrate, by measuring the fluctuation of activities within the \textit{Grid State}. Minimal variation across generations suggests a lack of significant activity within the system, while noticeable fluctuations indicate a more dynamic behaviour. Persistent and long-term fluctuations over time suggest long-term dynamics within the cellular automaton's \textit{Grid State}. 



Additionally, the count of cells in different \textit{Cell States} is tracked for each generation, serving as a measurement of the distribution of cells across various states (alive, decay, quiescent), independent of the actual computation happening on the \textit{Grid State}. 




\begin{table}[t]
  \centering
  \caption{Qualitative Phenotypic and Genotypic Measures}
  \label{tab:qual_measures}
  \begin{tabular}{cc}
    \toprule
    \textbf{Measure} & \textbf{Grid representation} \\
    \midrule
    Genotypic variability & Color-coded genome \\
     & \includegraphics[width=0.185\textwidth]{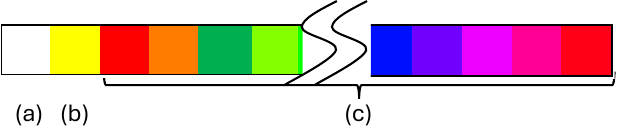}\\
    \textit{Cell state} and age & Tri-color (color shades for age)\\
     & \includegraphics[width=0.185\textwidth]{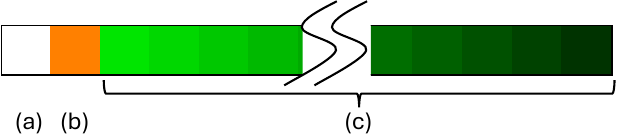}\\
    \textit{Grid state} & Tri-color (quiescent in white) \\
     & \includegraphics[width=0.05\textwidth]{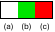}\\
    \bottomrule
  \end{tabular}
\vspace{-10pt}
\end{table}

\begin{table}[t]
  \centering
  \caption{Quantitative Phenotypic and Genotypic Measures}
  \label{tab:quant_measures}
  \begin{tabular}{cc}
    \toprule
    \textbf{Measure} \\
    \midrule
    Cumulative number of discovered rules \\
    \textit{Grid state} fluctuation  \\
    Number of alive, decay, and quiescent cells  \\
    Number of cells in state $0$, $1$, and quiescent\\
    \bottomrule
  \end{tabular}
\vspace{-10pt}
\end{table}

\section{Experimental Setup}

We conduct a series of experiments to identify suitable conditions that may support long-term dynamics in our heterogeneous life-like CA with age constrains and local evolution. In this section, we provide details on the experimental setup and utilized parameters. 



\begin{table}[t]
  \centering
  \caption{CA Parameters}
  \label{tab:parameters_ca}
  \begin{tabular}{cc}
    \toprule
    \textbf{Parameter} & \textbf{Value} \\
    \midrule
    Eligible Cell State to Inherit Genomes & [Alive] \\
    Probability Initial Alive Cell State & 0.5 \\
    Probability Alive Cell State with Value 1  & 0.5 \\
    Inheritance Probability  & 0.125 \\
    Mutation Probability $P_{mut}$ & 0.2 \\
    $a_{max}$ & 10, 50 \\
    $a_{dec}$ & 15, 70 \\    
    Initial Rule & B3S23 \\
    \bottomrule
  \end{tabular}
\vspace{-10pt}
\end{table}

The used parameters are given in Table \ref{tab:parameters_ca}. In particular, the \textit{Probability Initial Alive Cell State} determines the ratio of alive to quiescent cells in the first generation of the CA. In our experiments, the initial generation of a CA begins with expectedly half of the cells in a quiescent state, while the remaining half are alive. Initially, all alive cells start with a GoL genome, i.e., B3S23. Among the living cells, a stochastic allocation designates half with a phenotypic state of 0, while the remaining half is assigned a phenotypic state of 1, as dictated by the parameter \textit{Probability Alive Cell State with Value 1}. The \textit{Inheritance Probability}, set to 0.125, governs the likelihood of quiescent cells transitioning to an alive state, i.e. on average one out of the eight neighbors to a single alive cell is expected to become alive. Only alive cells are eligible for genome transmission to quiescent cells, determined by the parameter \textit{Eligible Cell State to Inherit Genomes}. This means that decay cells cannot reproduce. The inherited genomes of new alive cells may undergo mutation with a probability indicated by \textit{Mutation Probability ($P_{mut}$)}. Suitable values for \textit{Inheritance Probability} and \textit{Mutation Probability} used in this study are determined experimentally. 

Additionally, we experiment with two age/decay budgets, ${a}_{max}$ and ${a}_{dec}$, namely 10/15 and 50/70.

Finally, we benchmark our results with two GoL versions:

\begin{itemize}

\item a GoL version with age constrains where $P_{mut}$ = 0.0, however $a_{max}$ = 50, $a_{dec}$ = 70. All the other parameters are kept as in Table \ref{tab:parameters_ca}

\item a classical GoL execution (no age and no mutation), where $P_{mut}$ = 0.0 and $a_{max}$ / $a_{dec}$ are set to a value larger than the number of executed generations. In this case, all cells are alive in the initial state, either in Cell State 0 or 1 (with 0.5 probability).  

\end{itemize}

The experiments are run on a 50 $\times$ 50 grid for 10,000 generations and on a 500 $\times$ 500 grid for 1,000 steps. The experiments were repeated 10 times and for the quantitative metrics, results were averaged and the deviation calculated. Sample videos are available here: \url{https://tinyurl.com/vrkfs63a}.

\section{Results and Discussion}



\begin{figure*}
    \centering
    \begin{subfigure}[t]{0.45\textwidth}
        \centering
        \includegraphics[width=\textwidth]{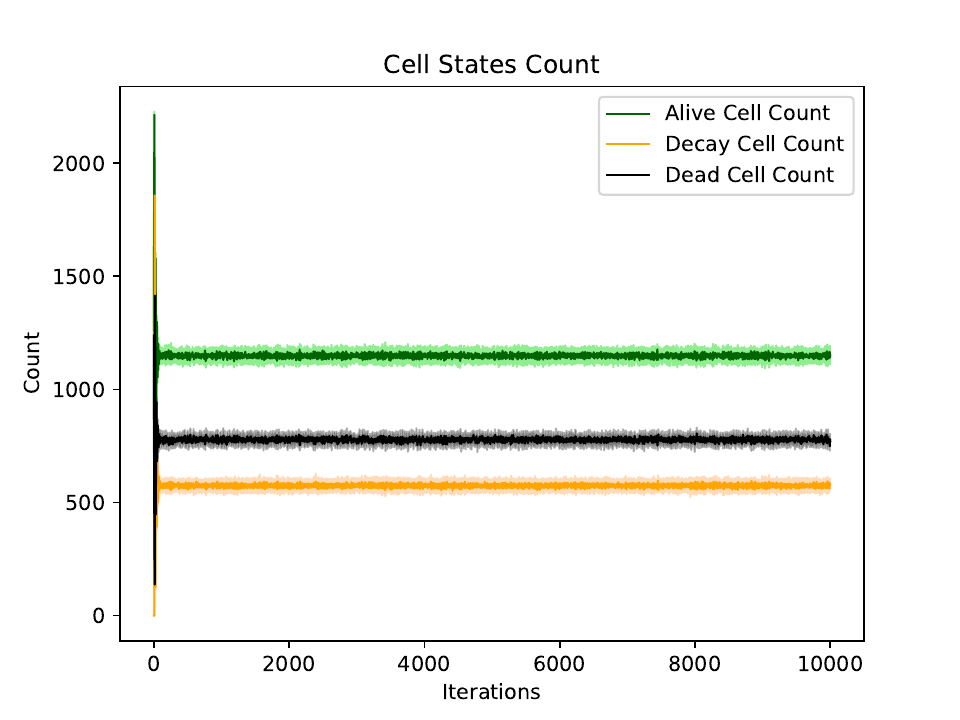}
        \caption{\textit{Cell States} count, $a_{max}$ 10 and $a_{dec}$ 15.}
        \label{fig:50x50_10K_cell_states_counts_amax10_adec15}
    \end{subfigure}
    \hfill
    \begin{subfigure}[t]{0.45\textwidth}
        \centering
        \includegraphics[width=\textwidth]{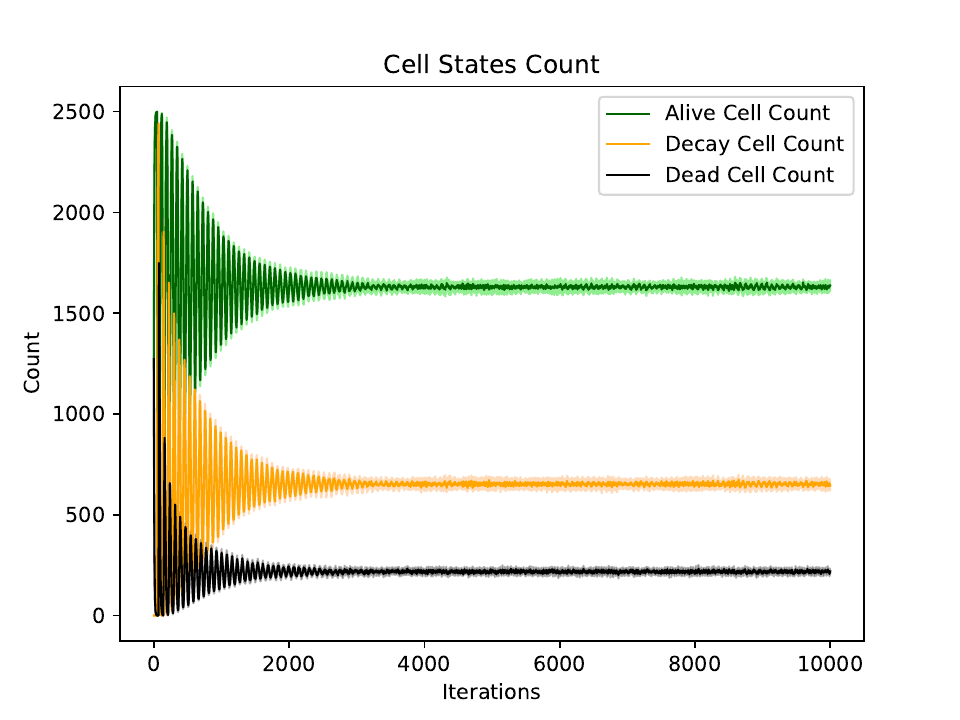}
        \caption{\textit{Cell States} count, $a_{max}$ 50 and $a_{dec}$ 70.}
        \label{fig:50x50_10K_cell_states_counts_amax50_adec70}
    \end{subfigure}
    \hfill
     \begin{subfigure}{0.45\textwidth}
        \centering
        \includegraphics[width=\textwidth]{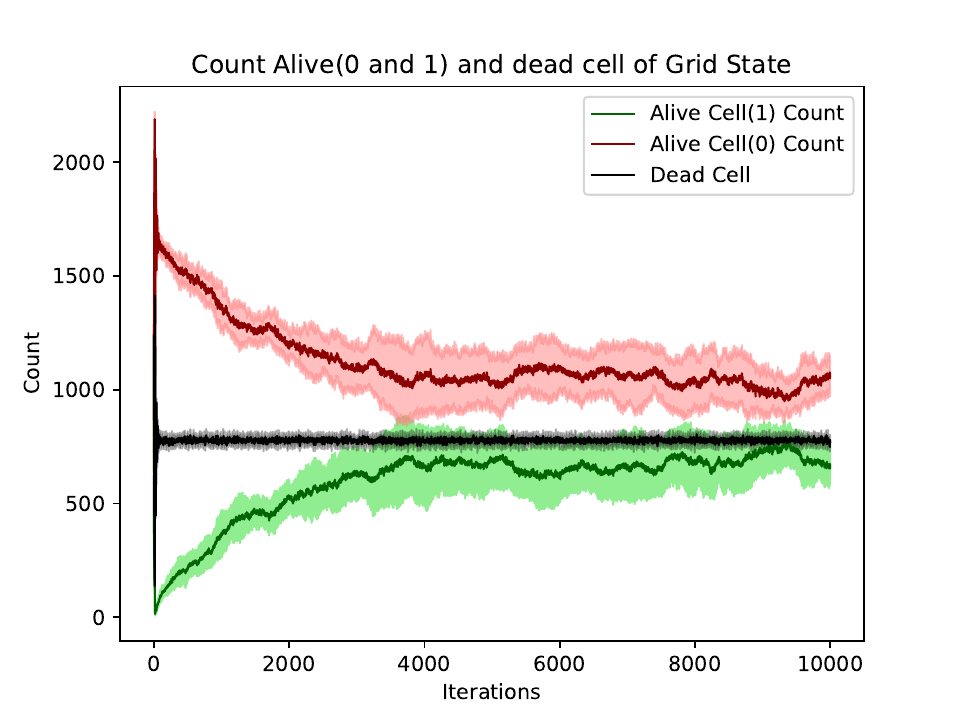}
        \caption{\textit{Grid States} count, $a_{max}$ 10 and $a_{dec}$ 15.}
        \label{fig:50x50_10K_grid_states_counts_amax10_adec15}
    \end{subfigure}
    \hfill
     \begin{subfigure}{0.45\textwidth}
        \centering
        \includegraphics[width=\textwidth]{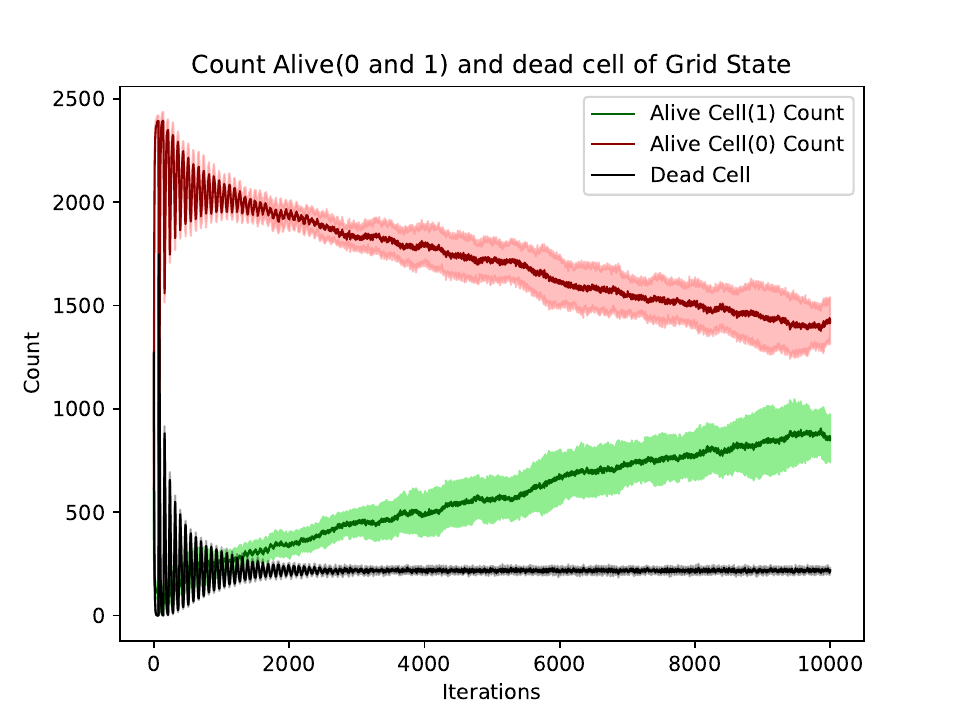}
        \caption{\textit{Grid States} count, $a_{max}$ 50 and $a_{dec}$ 70.}
        \label{fig:50x50_10K_grid_states_counts_amax50_adec70}
    \end{subfigure}
    \hspace*{-0.7cm}
    \hfill
     \begin{subfigure}{0.68\textwidth}
        \centering
        \includegraphics[height= 5.5cm, width=\textwidth]{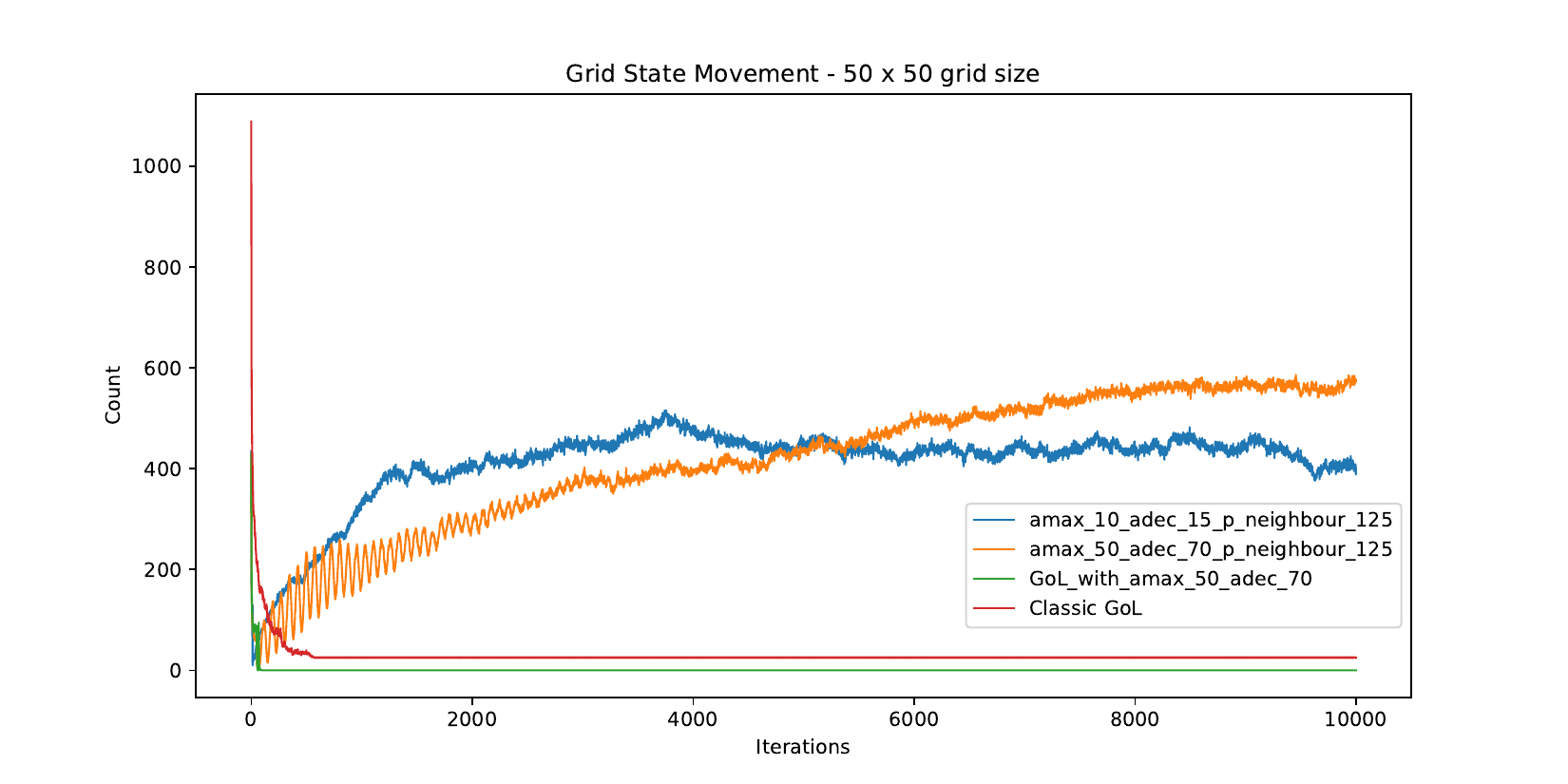}
        \caption{Comparison of \textit{Grid State} fluctuations.}
        \label{fig:50x50_10K_grid_state_movement}
    \end{subfigure}
    \hfill
    \begin{subfigure}{0.35\textwidth}
        \centering
        \includegraphics[height= 5.5cm,  width=\textwidth]{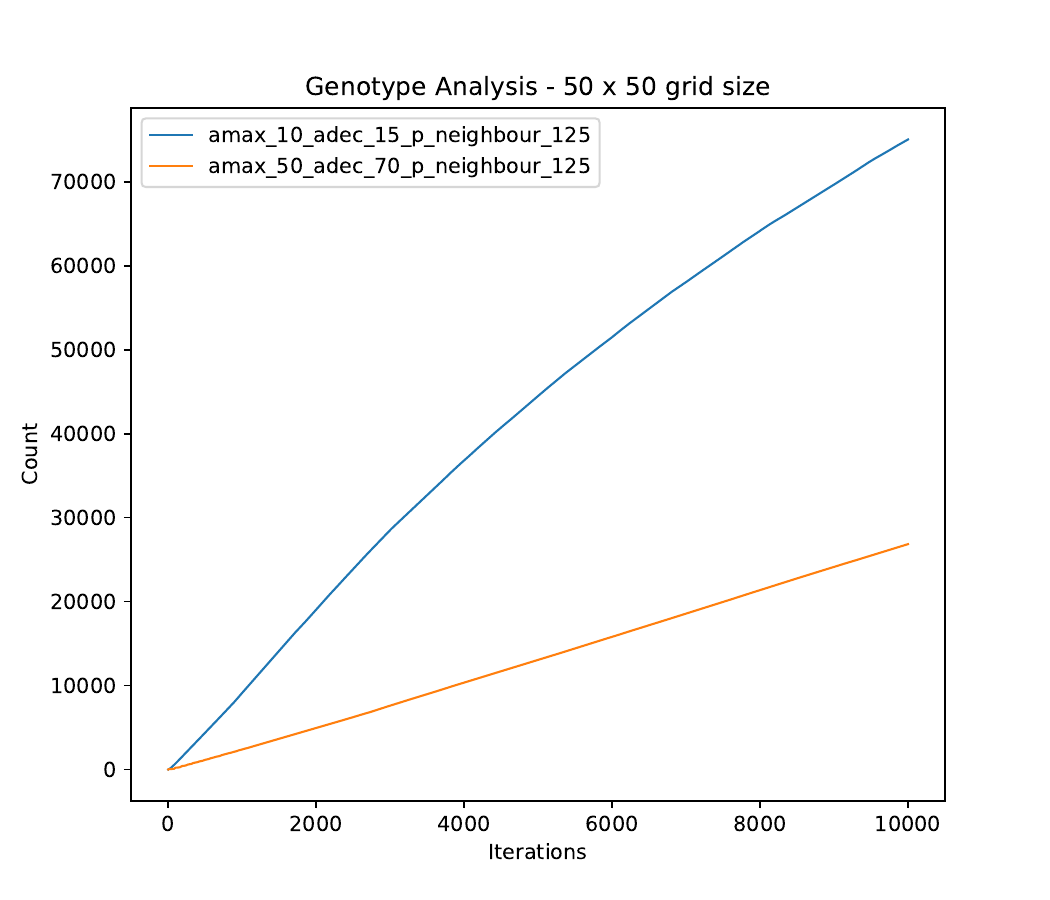}
        \caption{Cumulative Genotype count.}
        \label{fig:50x50_10K_genetic_analysis}
    \end{subfigure}

    \caption{Grid Size of 50 $\times$ 50 for 10,000 generations. Averages over 10 runs.}
    \label{fig:50x50gridsize_10kgen}
\end{figure*}

\begin{figure*}
    \centering
    \begin{subfigure}{0.45\textwidth}
        \centering
        \includegraphics[width=\textwidth]{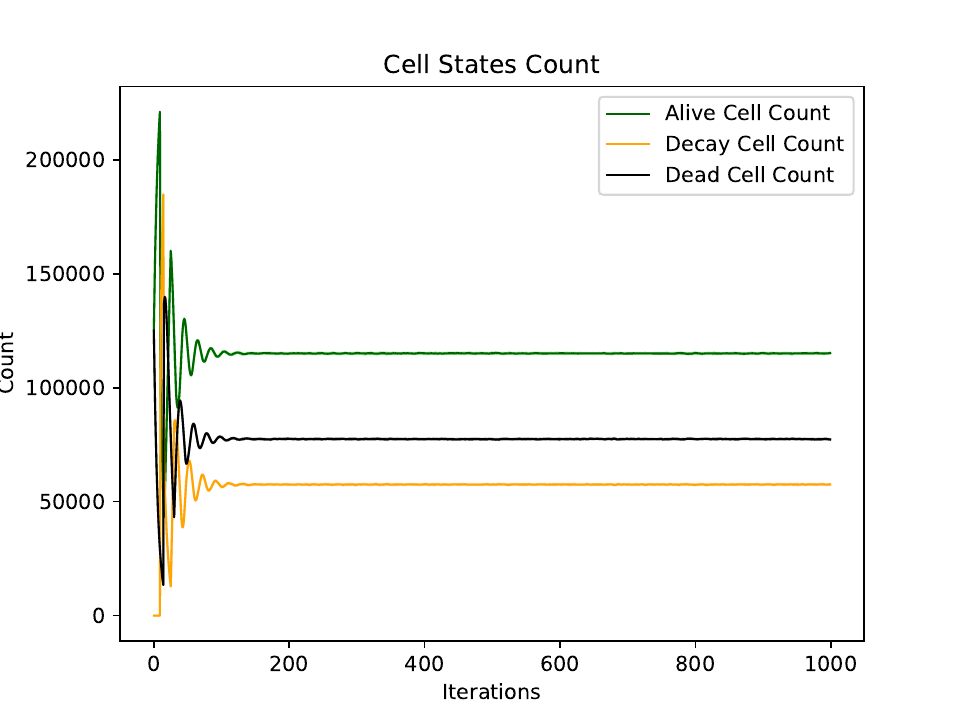}
        \caption{\textit{Cell States} count, $a_{max}$ 10 and $a_{dec}$ 15.}
        \label{fig:500x500_1K_cell_states_counts_amax10_adec15}
    \end{subfigure}
    \hfill
    \begin{subfigure}{0.45\textwidth}
        \centering
        \includegraphics[width=\textwidth]{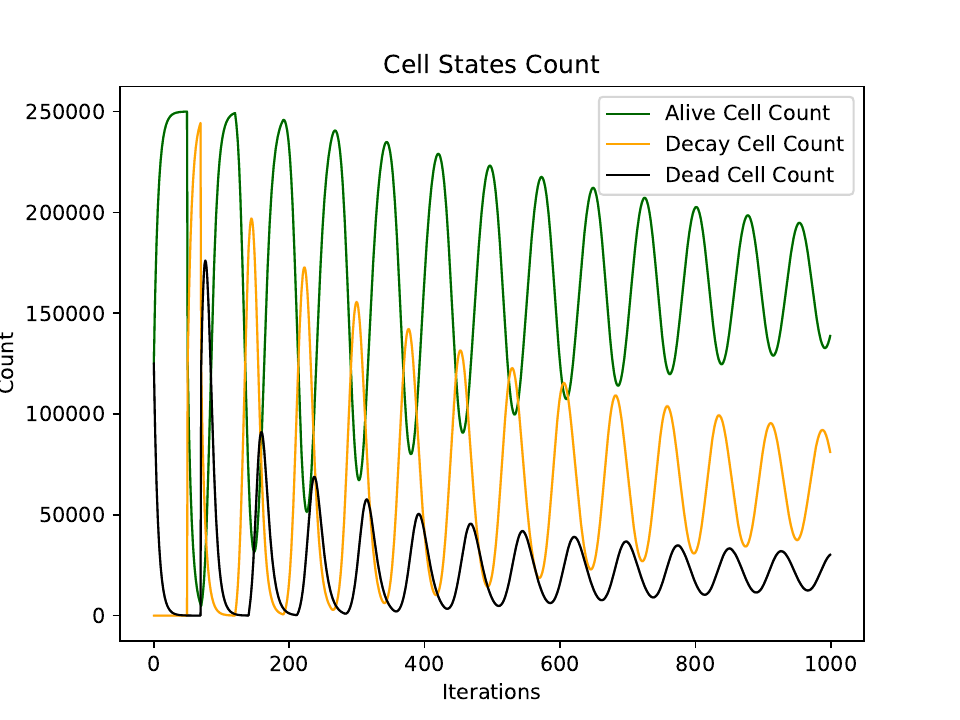}
        \caption{\textit{Cell States} Count, $a_{max}$ 50 and $a_{dec}$ 70.}
        \label{fig:500x500_1K_cell_states_counts_amax50_adec70}
    \end{subfigure}
    \hfill
     \begin{subfigure}{0.45\textwidth}
        \centering
        \includegraphics[width=\textwidth]{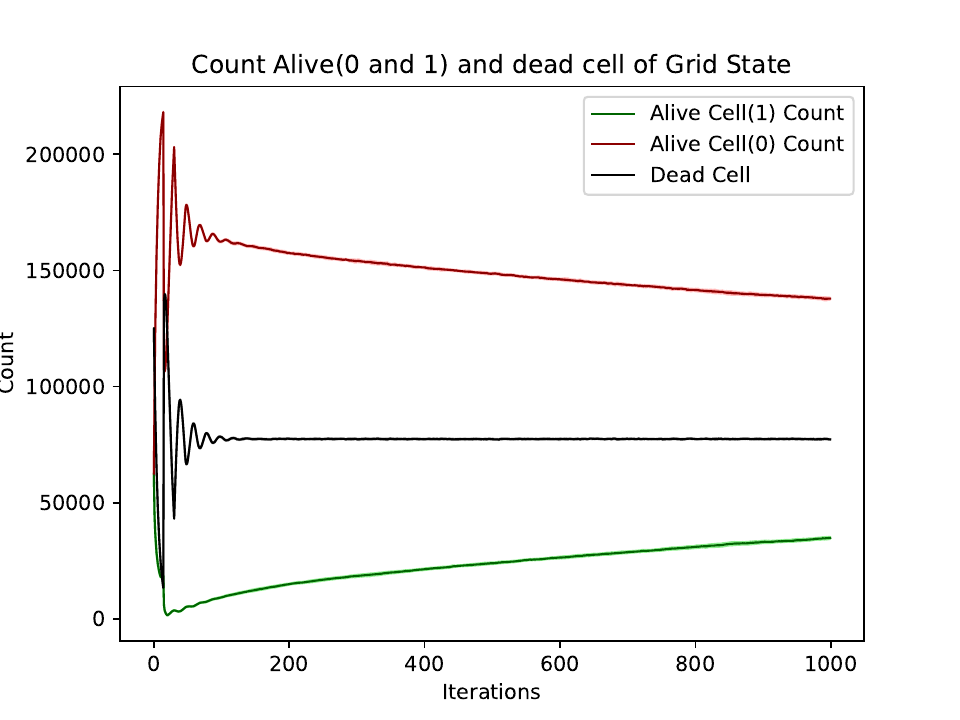}
        \caption{\textit{Grid States} count, $a_{max}$ 10 and $a_{dec}$ 15.}
        \label{fig:500x500_1K_grid_states_counts_amax10_adec15}
    \end{subfigure}
    \hfill
     \begin{subfigure}{0.45\textwidth}
        \centering
        \includegraphics[width=\textwidth]{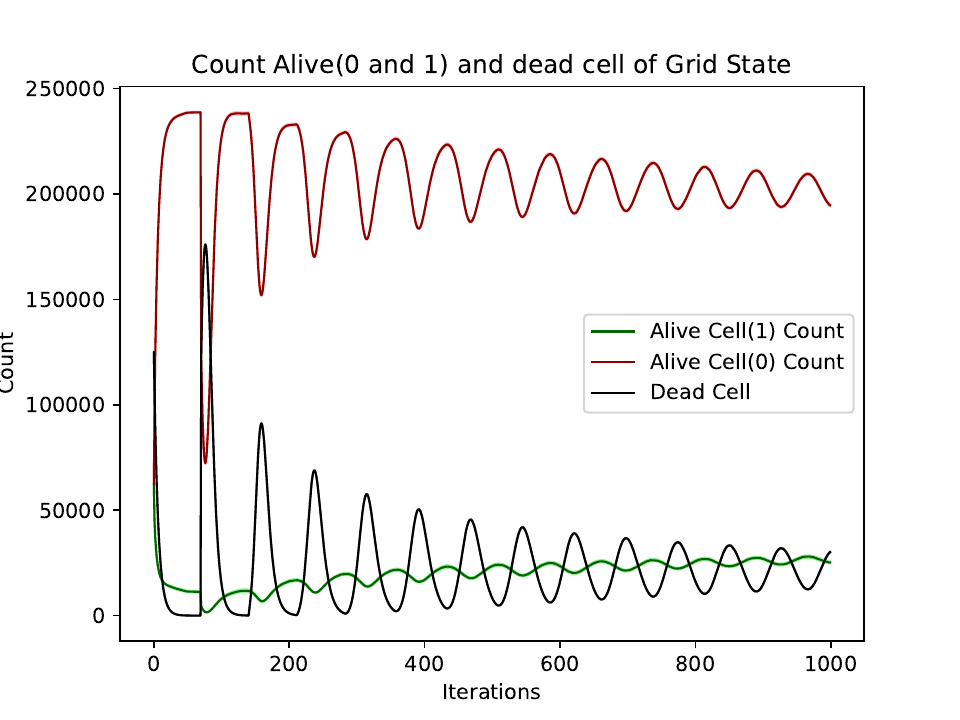}
        \caption{\textit{Grid States} count, $a_{max}$ 50 and $a_{dec}$ 70.}
        \label{fig:500x500_1K_grid_states_counts_amax50_adec70}
    \end{subfigure}
    \hfill
     \begin{subfigure}{0.45\textwidth}
        \centering
        \includegraphics[height= 5.5cm, width=\textwidth]{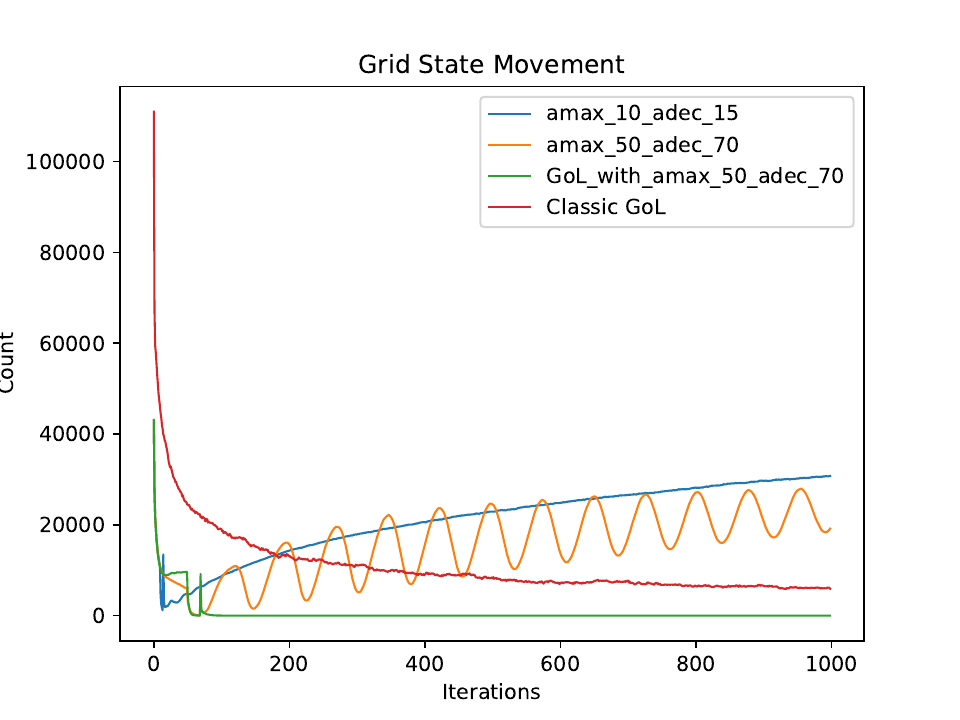}
        \caption{Comparison of \textit{Grid State} fluctuations.}
        \label{fig:500x500_1K_grid_state_movement}
    \end{subfigure}
    \hfill
    \begin{subfigure}{0.45\textwidth}
        \centering
        \includegraphics[height= 5.5cm,  width=\textwidth]{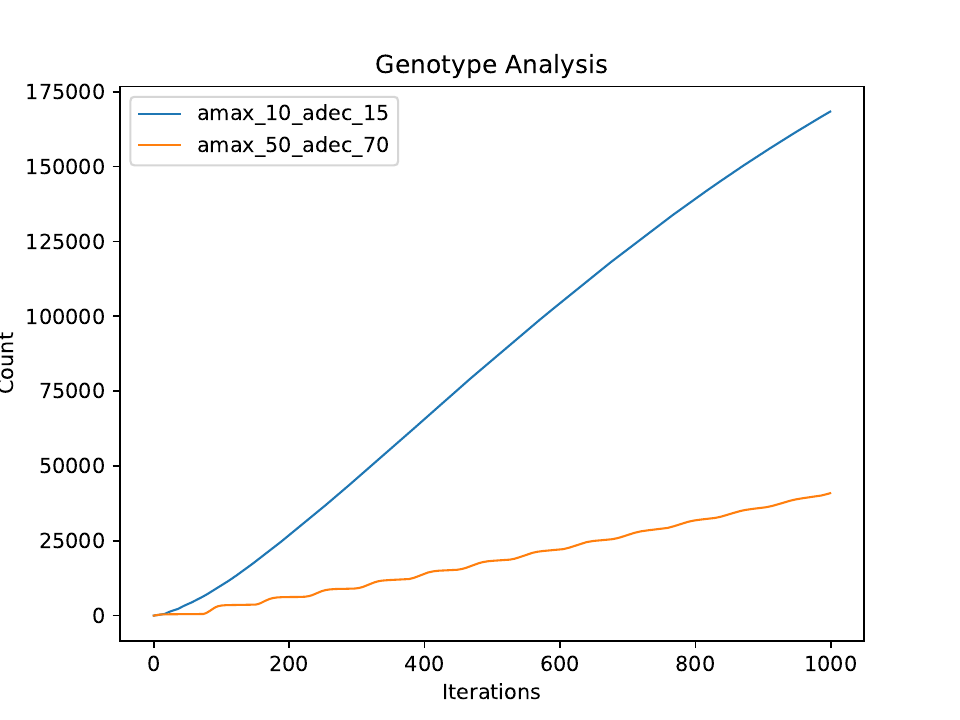}
        \caption{Cumulative Genotype count}
        \label{fig:500x500_1K_genetic_analysis}
    \end{subfigure}

    \caption{Grid Size of 500 $\times$ 500 for 1,000 generations. Averages over 10 runs.}
    \label{fig:500x500gridsize_1kgen}
\end{figure*}

The results for the experiments on the 50 $\times$ 50 grid are depicted in Figure \ref{fig:50x50gridsize_10kgen} and for the larger grid size of  500 $\times$ 500 in Figure \ref{fig:500x500gridsize_1kgen}.
In \ref{fig:50x50_10K_cell_states_counts_amax10_adec15} and \ref{fig:50x50_10K_grid_states_counts_amax10_adec15} are the results for the shorter lifespan of ${a}_{max} = 10$ and ${a}_{dec} = 15$. Here, all counts have large variability for the first few generations. After this initial period, the \textit{Cell State} counts seem to be almost constant with small deviations afterwards. In this equilibrium there are consistently more alive than dead and more dead than decaying cells. Considering the choice for ${a}_{max}$ and ${a}_{dec}$, the ratio of alive to decaying cells seems natural in such a state. Whether the amount of dead cells can be simply explained by the inheritance and mutation probability needs to be investigated.
For the \textit{Grid States}, the count of dead cells is the same as in the other plot. The amount of cells with state $0$ and $1$ seem to slowly converge, which means that the alive cell with state $0$ slowly become fewer and the ones with state $1$ increase for roughly the first 4,000 iterations. There is quite some variation, both for the mean as well as the deviations between runs. The growing amount of cells with state $1$ and its perseverance over 10,000 generations gives an indication of long-term phenotypic dynamics without signs of stagnation. 
For the shorter lifespan on the larger grid, \ref{fig:500x500_1K_cell_states_counts_amax10_adec15} and \ref{fig:500x500_1K_grid_states_counts_amax10_adec15} show similar behaviour as in the previously described results. After initial oscillation, the \textit{Cell State} counts stabilizes. The amount of cells with state $1$ and $0$ converge over time.

As can be seen in \ref{fig:50x50_10K_cell_states_counts_amax50_adec70} and \ref{fig:50x50_10K_grid_states_counts_amax50_adec70} of \ref{fig:50x50gridsize_10kgen}, for the longer lifespan of ${a}_{max} = 50$ and ${a}_{dec} = 70$, the \textit{Cell State} counts oscillate quite substantially in the beginning, but the oscillation's amplitude decrease exponentially in the first roughly 3,000 generations. Afterwards, the behaviour is again almost constant, with only little inter and intra run deviation. Alive cells are more numerous than decaying cells, which in turn occur more often than dead cells. For the \textit{Grid State}, the plot for the dead cells is naturally the same. The count of alive cells with state $0$ oscillates for the first 2,000 generations, but there is only little variability for the alive cells with state $1$. The curves for the alive cells with different grid state seem to converge up until the 9,000 generations, interestingly however this tendency does not seem to continue. 
For the longer lifespan of ${a}_{max} = 50$ and ${a}_{dec} = 70$, results are in \ref{fig:500x500_1K_cell_states_counts_amax50_adec70} and \ref{fig:500x500_1K_grid_states_counts_amax50_adec70}. The oscillation continues over the full $1,000$ generations of the experiment for both the cell and \textit{Grid State} counts. Overall, the plots are again similar to corresponding generations of \ref{fig:50x50_10K_cell_states_counts_amax50_adec70} and \ref{fig:50x50_10K_grid_states_counts_amax50_adec70}. After 100 generations, the amount of alive cells with state $1$ and dead cells are still roughly of the same size. In those plots, it becomes quite clear that the oscillations have different amplitudes: the count of alive and decaying cells are oscillating in similar strength, but oppositional. The same is true for the count of alive cells with state $0$ and the count of dead cells, but the overall amplitude is smaller. For the alive cells with state $1$ there is only small oscillation.

In plot \ref{fig:50x50_10K_grid_state_movement}, the fluctuation of \textit{Cell States} are depicted. The two benchmarks of GoL-like implementations have an early spike of \textit{Grid State} movement, but go down to a low, constant value. Our implementation with shorter lifespan has a noisy curve that, from about the 1,500 generation, seems to slowly fluctuate around a value of 400. The \textit{Grid State} movement corresponding to the longer lifespan is unsurprisingly unstable in the first 2000 generations, before the oscillations fades. However, surprisingly the longer lifespan curve shows more phenotypic fluctuations than the shorter lifespan. Both show rather high dynamic behaviors of around 400 to 600 grid changes between consecutive generations.
For the larger grid and shorter duration, the fluctuation of \textit{Cell States} in \ref{fig:500x500_1K_grid_state_movement} increases over time for both our configurations. After 200 generations, the value is almost always higher for the configuration with a small lifespan than for those of the other configurations. In case of the longer lifespan, the fluctuations oscillate for the full 1000 generations as in the experiment before.

As can be seen in plot \ref{fig:50x50_10K_genetic_analysis} and \ref{fig:500x500_1K_genetic_analysis}, the genotypic diversity increases over time for both of our implementations in both experimental setups, and when extrapolated one would expect more unique genomes to occur.
The model with a shorter lifespan on the larger grid discovers the most unique genomes with almost 175,000 of the possible 262,144. On the smaller grid, 70,000 genomes are discovered, but in a time span ten times longer. Also, it seems that the curve here is slowly saturating. The implementation with the longer lifespan discovers just over 20,000 and 35,000 genomes in the two experiments respectively.

\begin{figure}[ht!]
    \centering
    
    \begin{subfigure}[b]{0.093\textwidth}
        \includegraphics[width=0.92\linewidth]{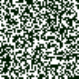}
        
        \label{fig:cellstates1}
    \end{subfigure}%
    \begin{subfigure}[b]{0.093\textwidth}
        \includegraphics[width=0.92\linewidth]{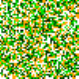}
        
        \label{fig:cellstates2500}
    \end{subfigure}%
    \begin{subfigure}[b]{0.093\textwidth}
        \includegraphics[width=0.92\linewidth]{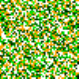}
        
        \label{fig:cellstates5000}
    \end{subfigure}%
    \begin{subfigure}[b]{0.093\textwidth}
        \includegraphics[width=0.92\linewidth]{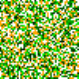}
        
        \label{fig:cellstates7500}
    \end{subfigure}%
    \begin{subfigure}[b]{0.093\textwidth}
        \includegraphics[width=0.92\linewidth]{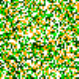}
        
        \label{fig:cellstates10000}
    \end{subfigure}

    \medskip

    \begin{subfigure}[b]{0.093\textwidth}
        \includegraphics[width=0.92\linewidth]{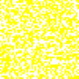}
        
        \label{fig:genomes1}
    \end{subfigure}%
    \begin{subfigure}[b]{0.093\textwidth}
        \includegraphics[width=0.92\linewidth]{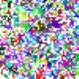}
        
        \label{fig:genomes2500}
    \end{subfigure}%
    \begin{subfigure}[b]{0.093\textwidth}
        \includegraphics[width=0.92\linewidth]{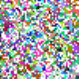}
        
        \label{fig:genomes5000}
    \end{subfigure}%
    \begin{subfigure}[b]{0.093\textwidth}
        \includegraphics[width=0.92\linewidth]{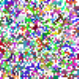}
        
        \label{fig:genomes7500}
    \end{subfigure}%
    \begin{subfigure}[b]{0.093\textwidth}
        \includegraphics[width=0.92\linewidth]{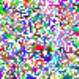}
        
        \label{fig:genomes10000}
    \end{subfigure}

    \medskip

    \begin{subfigure}[b]{0.093\textwidth}
        \includegraphics[width=0.92\linewidth]{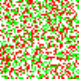}
        
        \label{fig:gridstates1}
    \end{subfigure}%
    \begin{subfigure}[b]{0.093\textwidth}
        \includegraphics[width=0.92\linewidth]{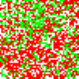}
        
        \label{fig:gridstates2500}
    \end{subfigure}%
    \begin{subfigure}[b]{0.093\textwidth}
        \includegraphics[width=0.92\linewidth]{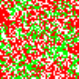}
        
        \label{fig:gridstates5000}
    \end{subfigure}%
    \begin{subfigure}[b]{0.093\textwidth}
        \includegraphics[width=0.92\linewidth]{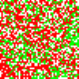}
        
        \label{fig:gridstates7500}
    \end{subfigure}%
    \begin{subfigure}[b]{0.093\textwidth}
        \includegraphics[width=0.92\linewidth]{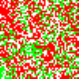}
        
        \label{fig:gridstates10000}
    \end{subfigure}

    \caption{Example execution of 50 $\times$ 50 grid, $a_{max}$ = 10, $a_{dec}$ = 15. First Row: \textit{Cell States}, Second Row: Genomes, and Third Row: \textit{Grid States} - Columns: snapshots at generation 1, 2,500, 5,000, 7,500, 10,000.}
    \label{fig:505010k_1015}
\end{figure}

\begin{figure}[ht!]
    \centering
    \begin{subfigure}[b]{0.093\textwidth}
        \includegraphics[width=0.92\linewidth]{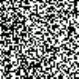}
        
        \label{fig:cellstates1}
    \end{subfigure}%
    \begin{subfigure}[b]{0.093\textwidth}
        \includegraphics[width=0.92\linewidth]{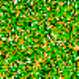}
        
        \label{fig:cellstates2500}
    \end{subfigure}%
    \begin{subfigure}[b]{0.093\textwidth}
        \includegraphics[width=0.92\linewidth]{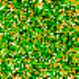}
        
        \label{fig:cellstates5000}
    \end{subfigure}%
    \begin{subfigure}[b]{0.093\textwidth}
        \includegraphics[width=0.92\linewidth]{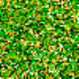}
        
        \label{fig:cellstates7500}
    \end{subfigure}%
    \begin{subfigure}[b]{0.093\textwidth}
        \includegraphics[width=0.92\linewidth]{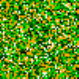}
        
        \label{fig:cellstates10000}
    \end{subfigure}

    \medskip

    \begin{subfigure}[b]{0.093\textwidth}
        \includegraphics[width=0.92\linewidth]{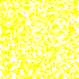}
        
        \label{fig:genomes1}
    \end{subfigure}%
    \begin{subfigure}[b]{0.093\textwidth}
        \includegraphics[width=0.92\linewidth]{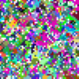}
        
        \label{fig:genomes2500}
    \end{subfigure}%
    \begin{subfigure}[b]{0.093\textwidth}
        \includegraphics[width=0.92\linewidth]{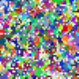}
        
        \label{fig:genomes5000}
    \end{subfigure}%
    \begin{subfigure}[b]{0.093\textwidth}
        \includegraphics[width=0.92\linewidth]{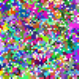}
        
        \label{fig:genomes7500}
    \end{subfigure}%
    \begin{subfigure}[b]{0.093\textwidth}
        \includegraphics[width=0.92\linewidth]{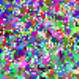}
        
        \label{fig:genomes10000}
    \end{subfigure}

    \medskip

    \begin{subfigure}[b]{0.093\textwidth}
        \includegraphics[width=0.92\linewidth]{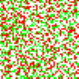}
        
        \label{fig:gridstates1}
    \end{subfigure}%
    \begin{subfigure}[b]{0.093\textwidth}
        \includegraphics[width=0.92\linewidth]{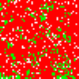}
        
        \label{fig:gridstates2500}
    \end{subfigure}%
    \begin{subfigure}[b]{0.093\textwidth}
        \includegraphics[width=0.92\linewidth]{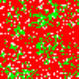}
        
        \label{fig:gridstates5000}
    \end{subfigure}%
    \begin{subfigure}[b]{0.093\textwidth}
        \includegraphics[width=0.92\linewidth]{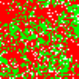}
        
        \label{fig:gridstates7500}
    \end{subfigure}%
    \begin{subfigure}[b]{0.093\textwidth}
        \includegraphics[width=0.92\linewidth]{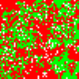}
        
        \label{fig:gridstates10000}
    \end{subfigure}

    \caption{Example execution of 50 $\times$ 50 grid, $a_{max}$ = 50, $a_{dec}$ = 70. First Row: \textit{Cell States}, Second Row: Genomes, and Third Row: \textit{Grid States} - Columns: snapshots at generation 1, 2,500, 5,000, 7,500, 10,000.}
    \label{fig:505010k_5070}
\end{figure}

\begin{figure*}[htbp]
        \centering
        \begin{subfigure}[b]{0.19\textwidth}
            \includegraphics[width=0.93\linewidth]{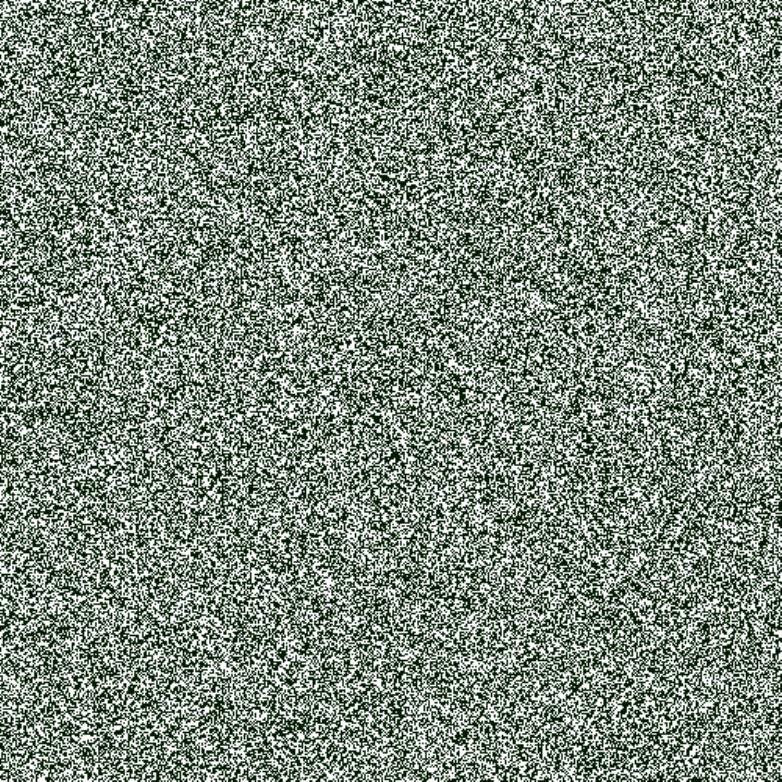}
            
            \label{fig:cellstates1}
        \end{subfigure}%
        \begin{subfigure}[b]{0.19\textwidth}
            \includegraphics[width=0.93\linewidth]{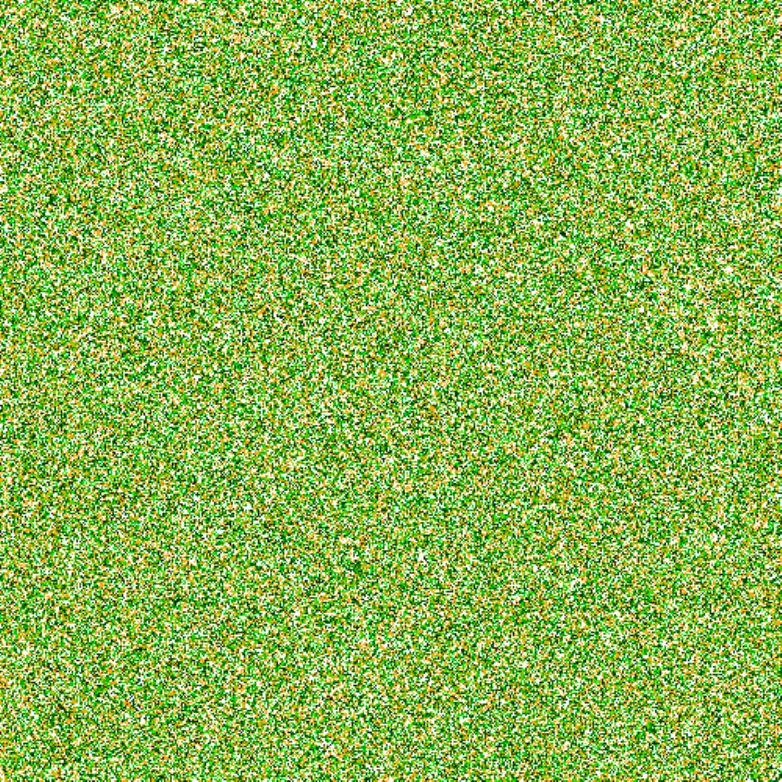}
            
            \label{fig:cellstates250}
        \end{subfigure}%
        \begin{subfigure}[b]{0.19\textwidth}
            \includegraphics[width=0.93\linewidth]{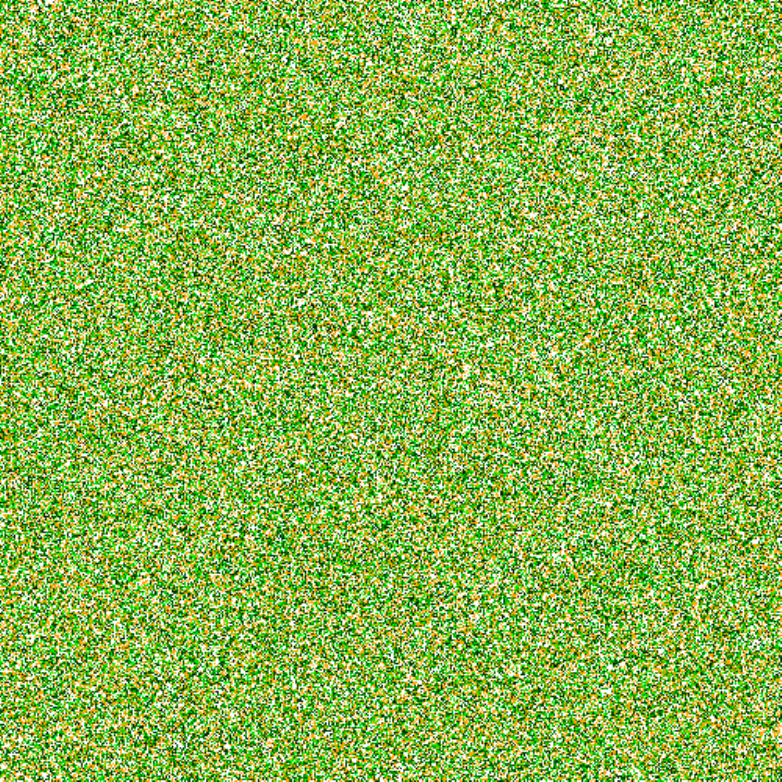}
            
            \label{fig:cellstates500}
        \end{subfigure}%
        \begin{subfigure}[b]{0.19\textwidth}
            \includegraphics[width=0.93\linewidth]{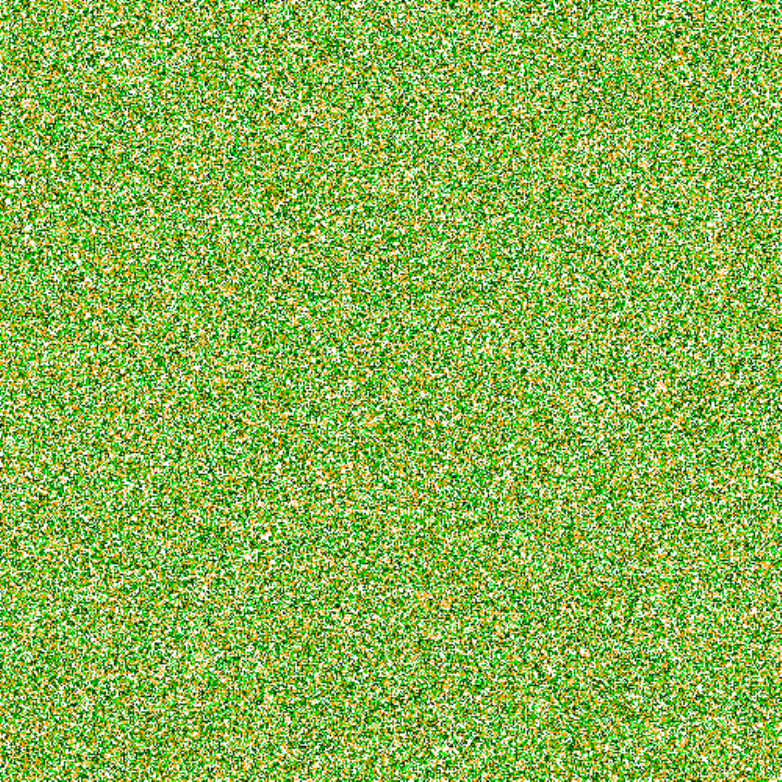}
            
            \label{fig:cellstates750}
        \end{subfigure}%
        \begin{subfigure}[b]{0.19\textwidth}
            \includegraphics[width=0.93\linewidth]{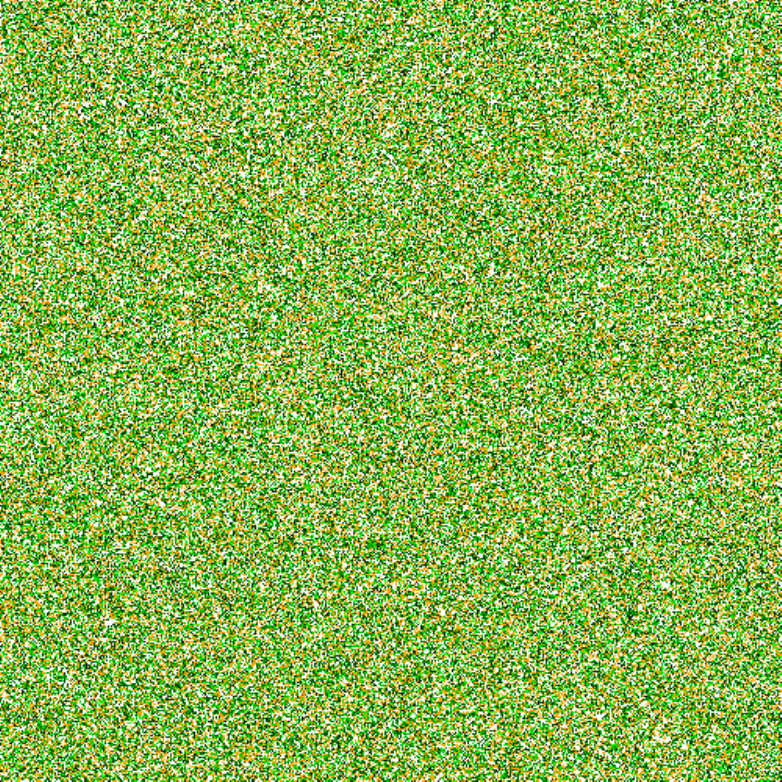}
            
            \label{fig:cellstates1000}
        \end{subfigure}
    
        \medskip
    
        \begin{subfigure}[b]{0.19\textwidth}
            \includegraphics[width=0.93\linewidth]{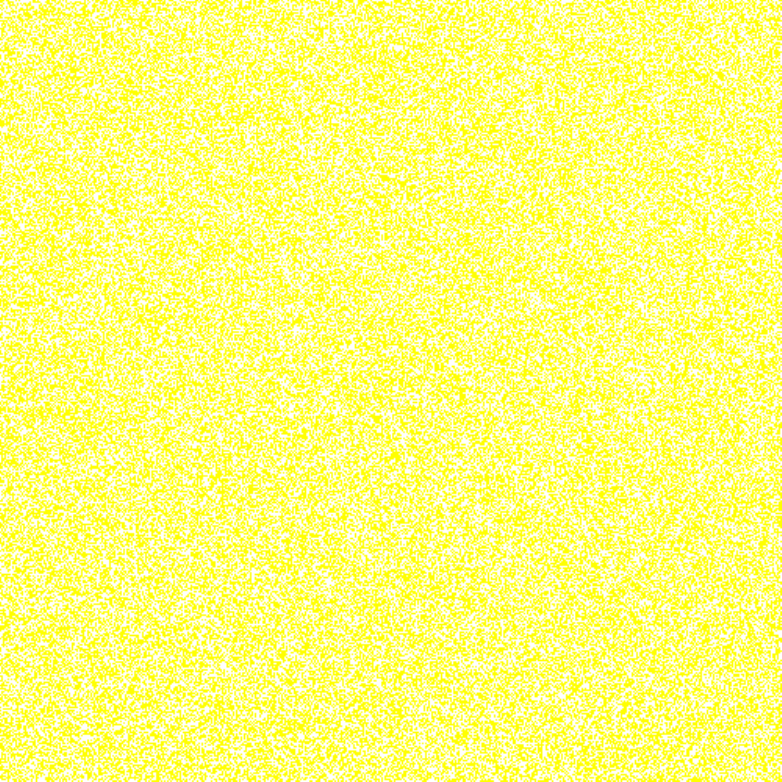}
            
            \label{fig:genomes1}
        \end{subfigure}%
        \begin{subfigure}[b]{0.19\textwidth}
            \includegraphics[width=0.93\linewidth]{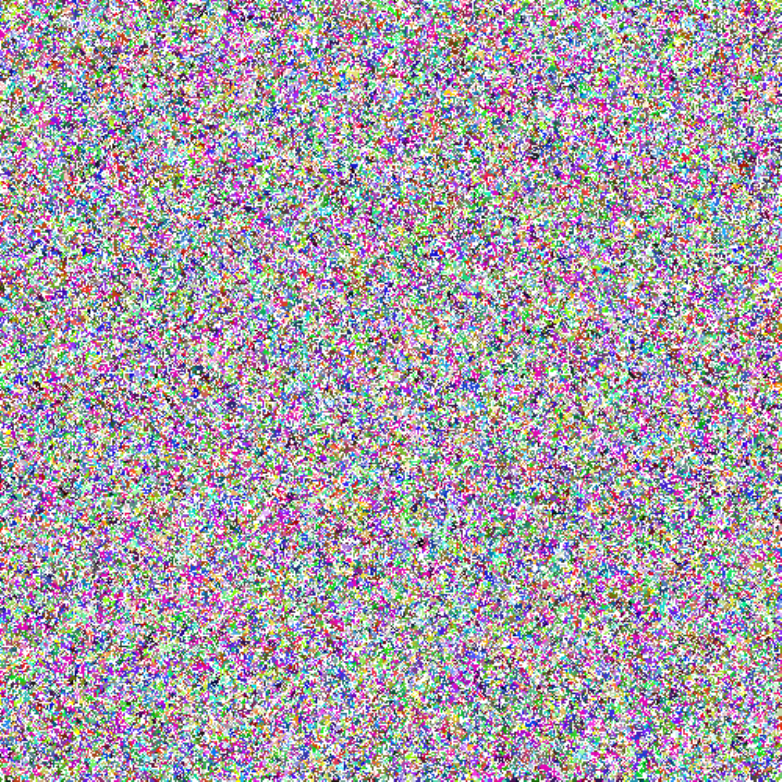}
            
            \label{fig:genomes250}
        \end{subfigure}%
        \begin{subfigure}[b]{0.19\textwidth}
            \includegraphics[width=0.93\linewidth]{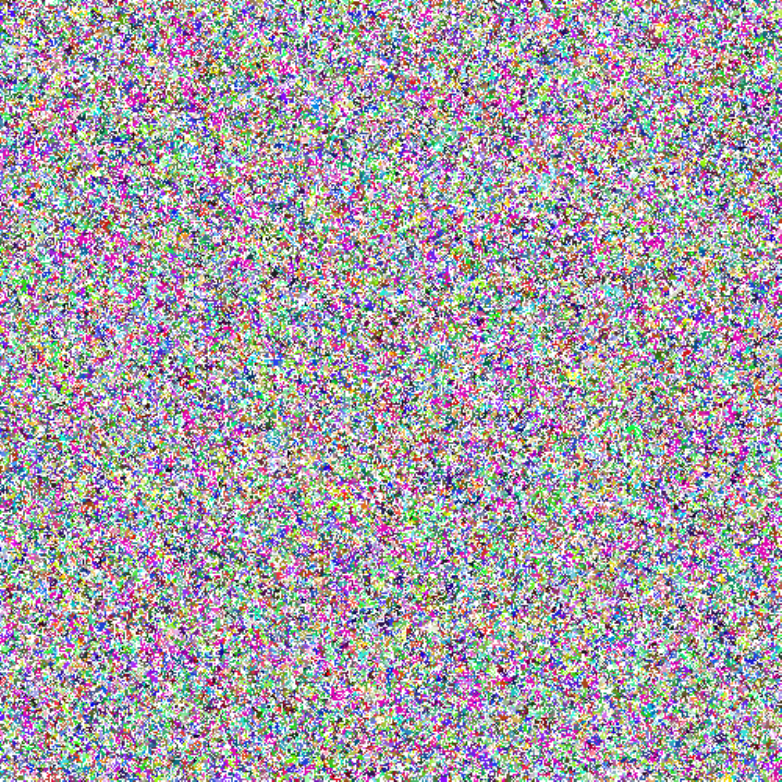}
            
            \label{fig:genomes500}
        \end{subfigure}%
        \begin{subfigure}[b]{0.19\textwidth}
            \includegraphics[width=0.93\linewidth]{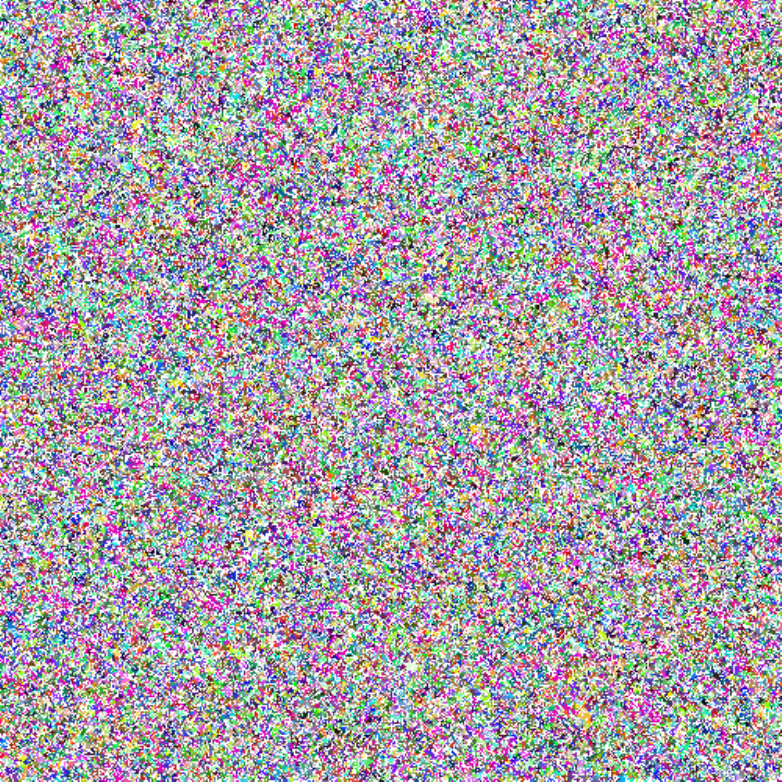}
            
            \label{fig:genomes750}
        \end{subfigure}%
        \begin{subfigure}[b]{0.19\textwidth}
            \includegraphics[width=0.93\linewidth]{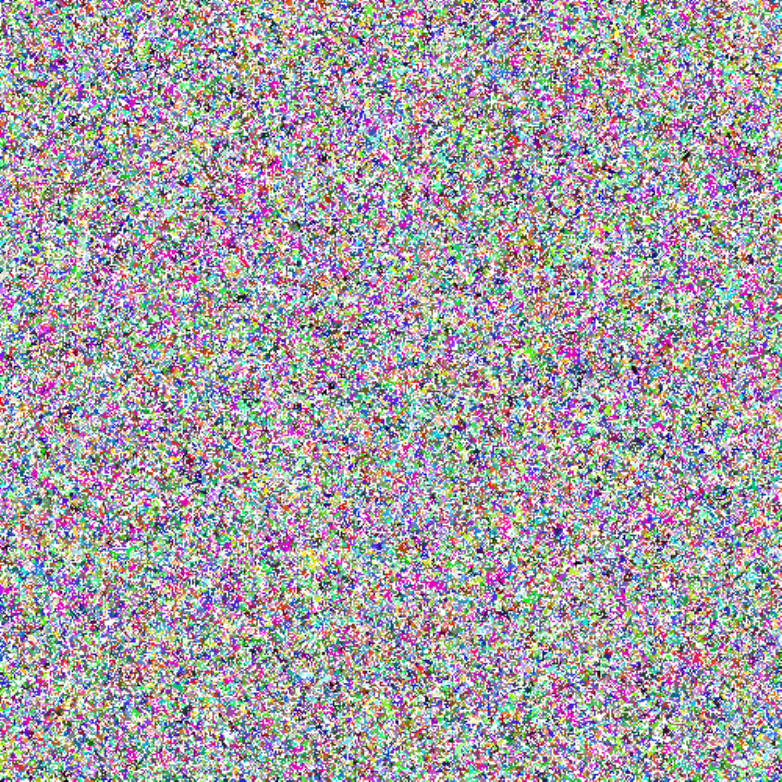}
            
            \label{fig:genomes1000}
        \end{subfigure}
    
        \medskip
    
        \begin{subfigure}[b]{0.19\textwidth}
            \includegraphics[width=0.93\linewidth]{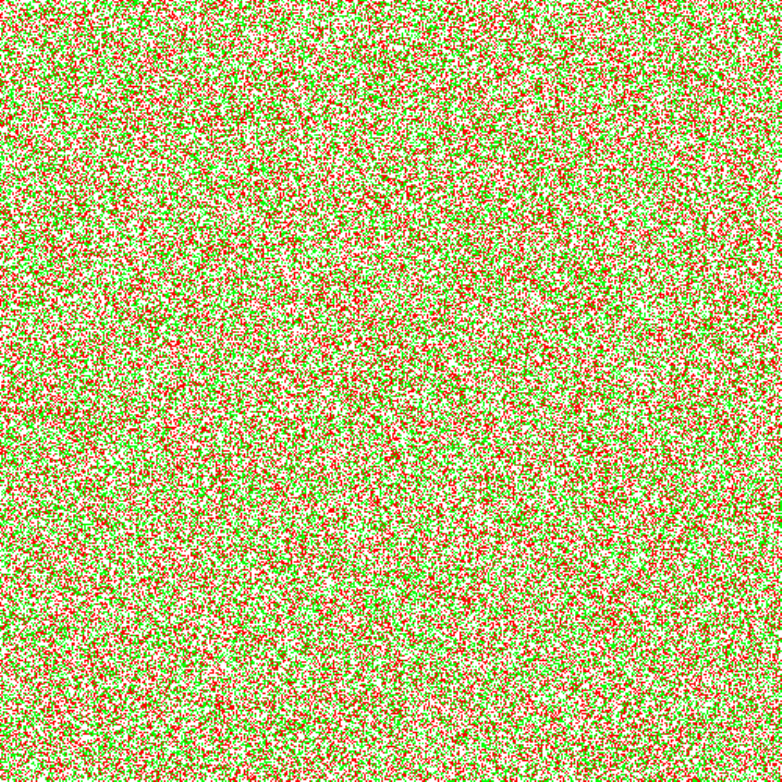}
            
            \label{fig:gridstates1}
        \end{subfigure}%
        \begin{subfigure}[b]{0.19\textwidth}
            \includegraphics[width=0.93\linewidth]{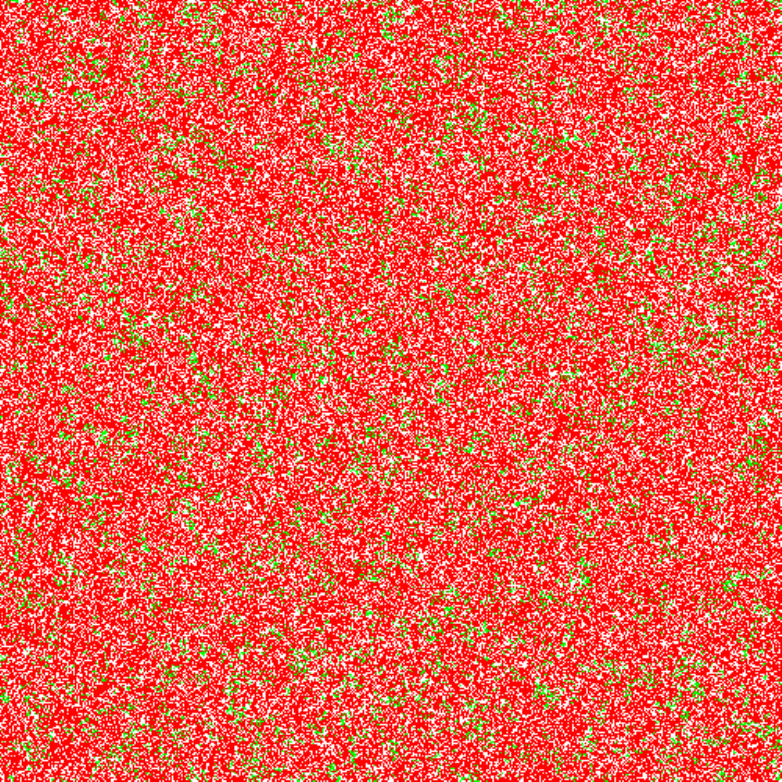}
            
            \label{fig:gridstates250}
        \end{subfigure}%
        \begin{subfigure}[b]{0.19\textwidth}
            \includegraphics[width=0.93\linewidth]{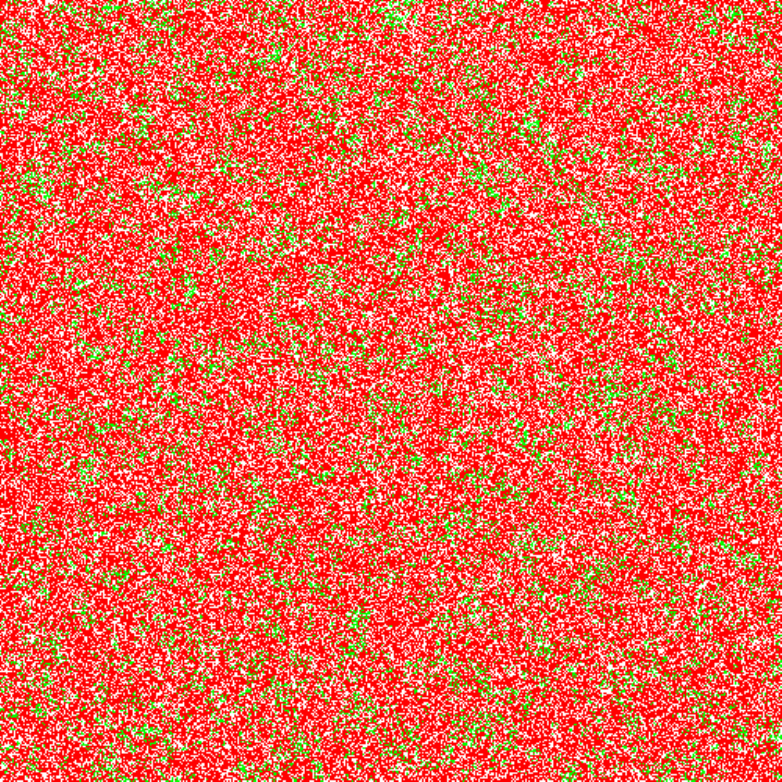}
            
            \label{fig:gridstates500}
        \end{subfigure}%
        \begin{subfigure}[b]{0.19\textwidth}
            \includegraphics[width=0.93\linewidth]{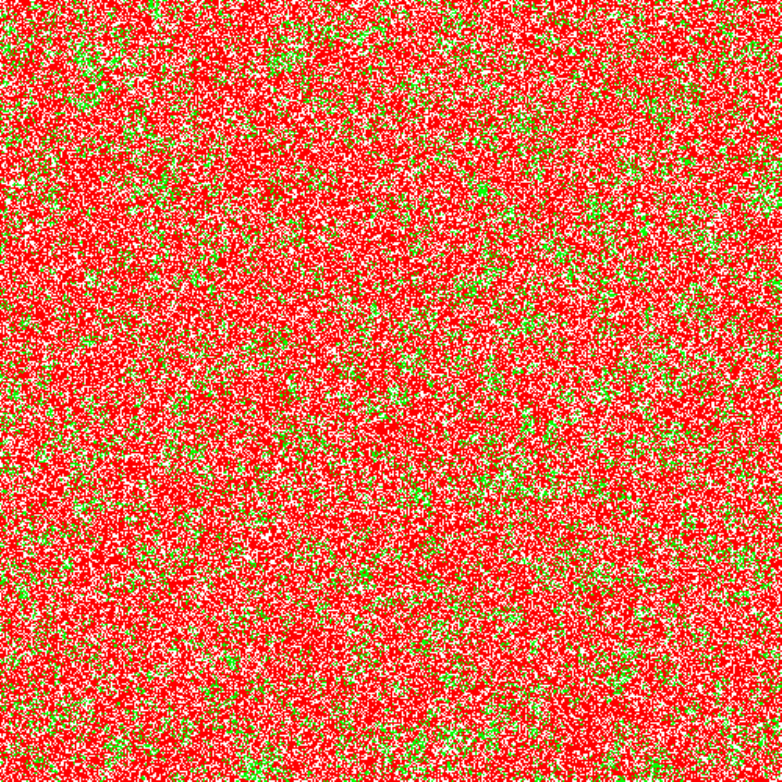}
            
            \label{fig:gridstates750}
        \end{subfigure}%
        \begin{subfigure}[b]{0.19\textwidth}
            \includegraphics[width=0.93\linewidth]{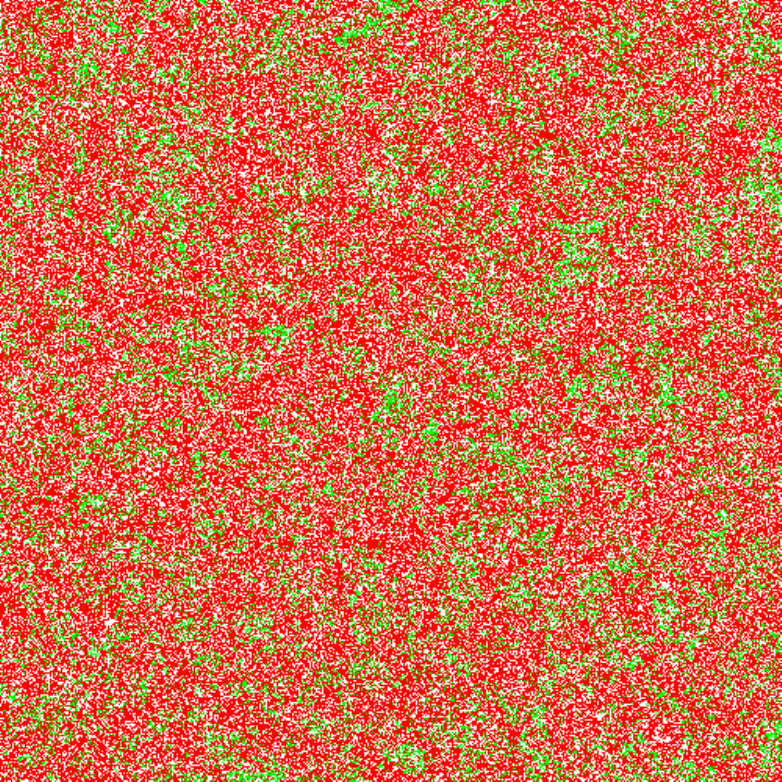}
            
            \label{fig:gridstates1000}
        \end{subfigure}
    
        \caption{Example execution of 500 $\times$ 500 grid, $a_{max}$ = 10, $a_{dec}$ = 15. First Row: \textit{Cell States}, Second Row: Genomes, and Third Row: \textit{Grid States} - Columns: Snapshots at generation 1, 250, 500, 750, 1,000.}
        \label{fig:5005001k_1015}
    \end{figure*}

\begin{figure*}[htbp]
        \centering
        \begin{subfigure}[b]{0.19\textwidth}
            \includegraphics[width=0.93\linewidth]{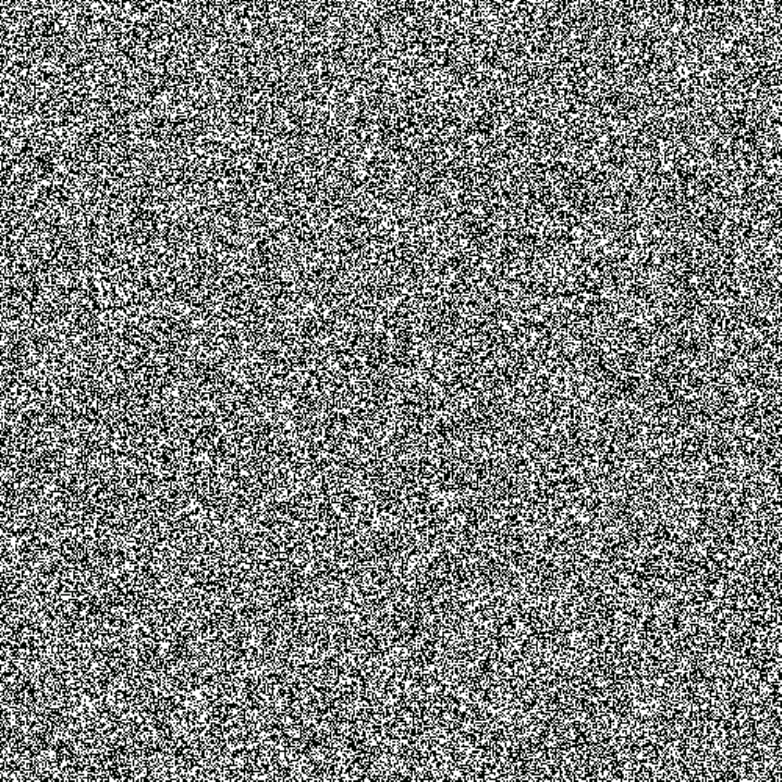}
            
            \label{fig:cellstates1}
        \end{subfigure}%
        \begin{subfigure}[b]{0.19\textwidth}
            \includegraphics[width=0.93\linewidth]{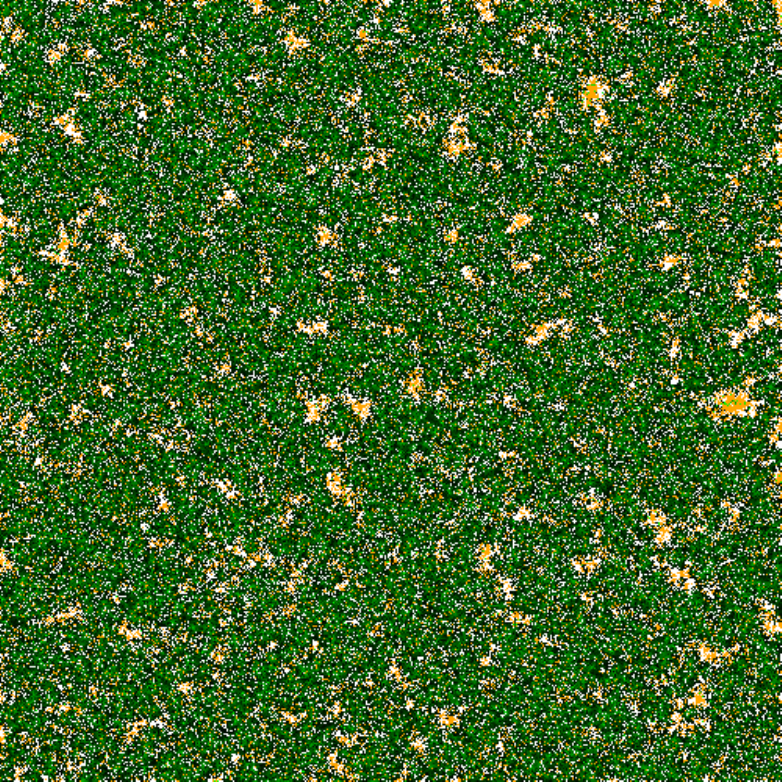}
            
            \label{fig:cellstates250}
        \end{subfigure}%
        \begin{subfigure}[b]{0.19\textwidth}
            \includegraphics[width=0.93\linewidth]{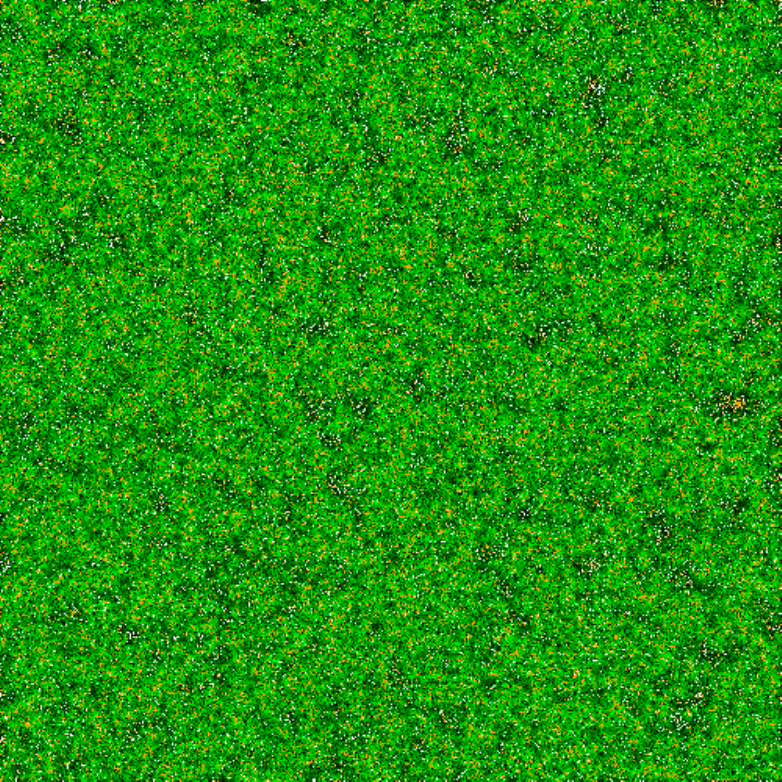}
            
            \label{fig:cellstates500}
        \end{subfigure}%
        \begin{subfigure}[b]{0.19\textwidth}
            \includegraphics[width=0.93\linewidth]{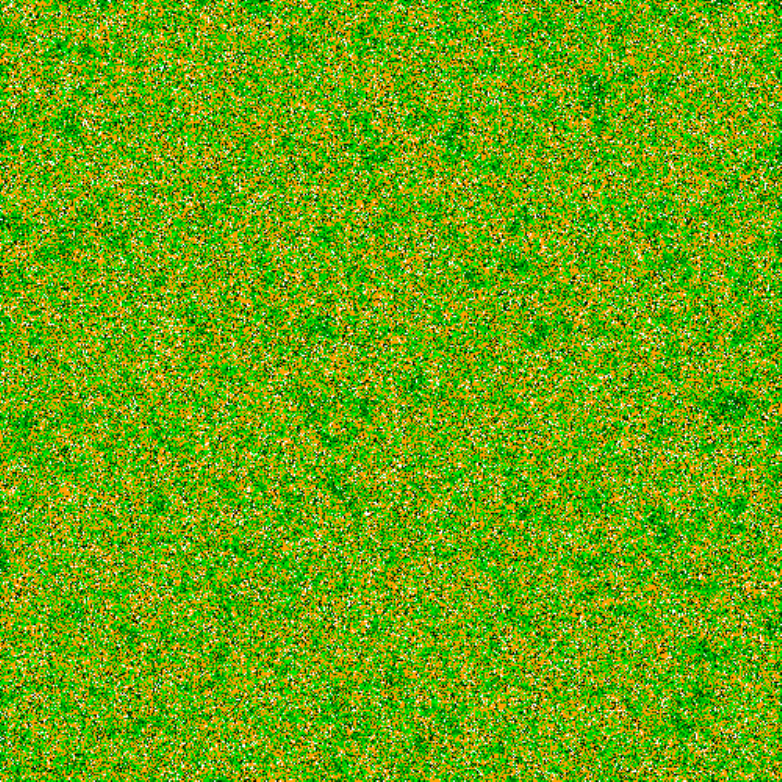}
            
            \label{fig:cellstates750}
        \end{subfigure}%
        \begin{subfigure}[b]{0.19\textwidth}
            \includegraphics[width=0.93\linewidth]{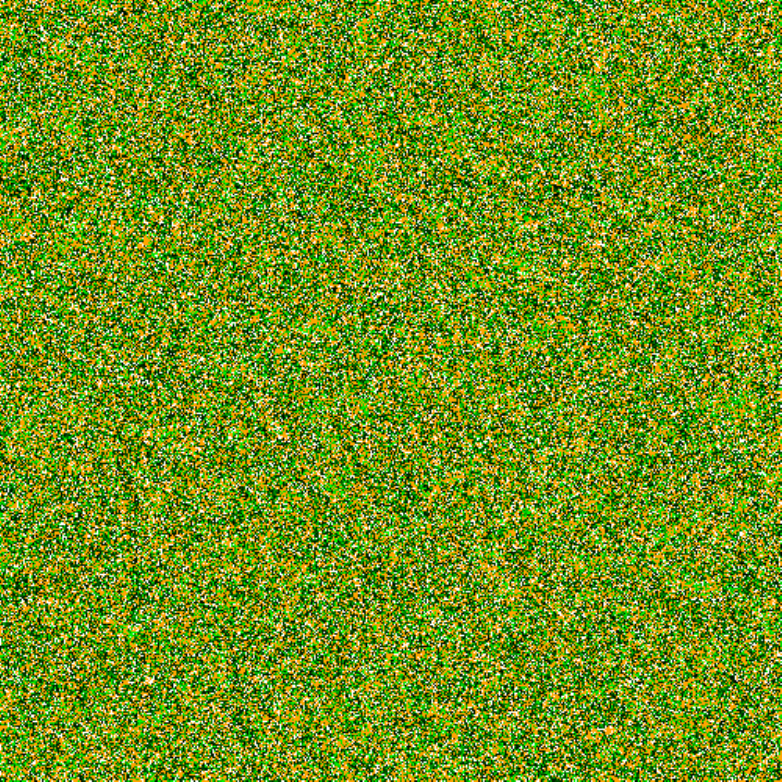}
            
            \label{fig:cellstates1000}
        \end{subfigure}
    
        \medskip
    
        \begin{subfigure}[b]{0.19\textwidth}
            \includegraphics[width=0.93\linewidth]{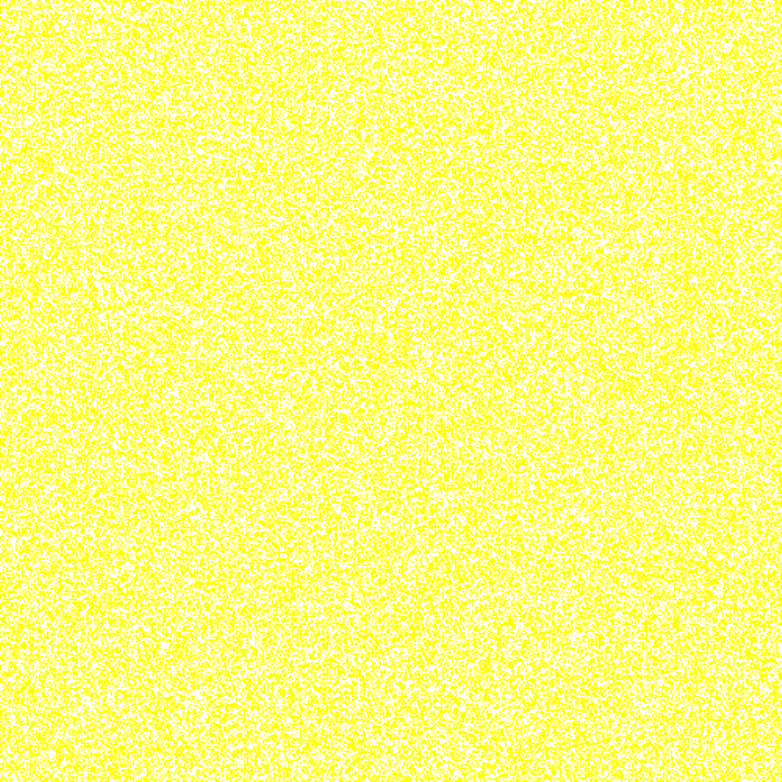}
            
            \label{fig:genomes1}
        \end{subfigure}%
        \begin{subfigure}[b]{0.19\textwidth}
            \includegraphics[width=0.93\linewidth]{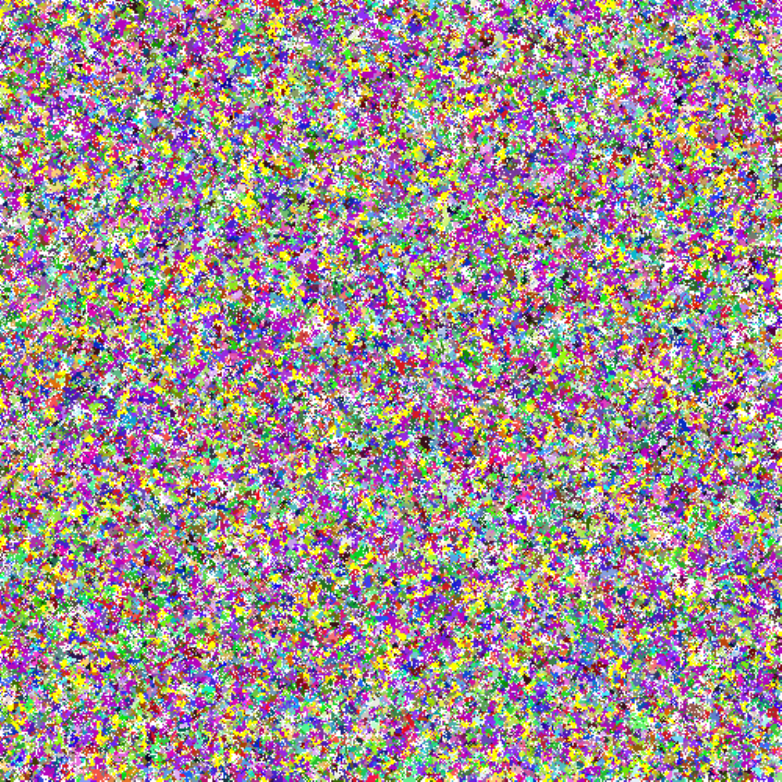}
            
            \label{fig:genomes250}
        \end{subfigure}%
        \begin{subfigure}[b]{0.19\textwidth}
            \includegraphics[width=0.93\linewidth]{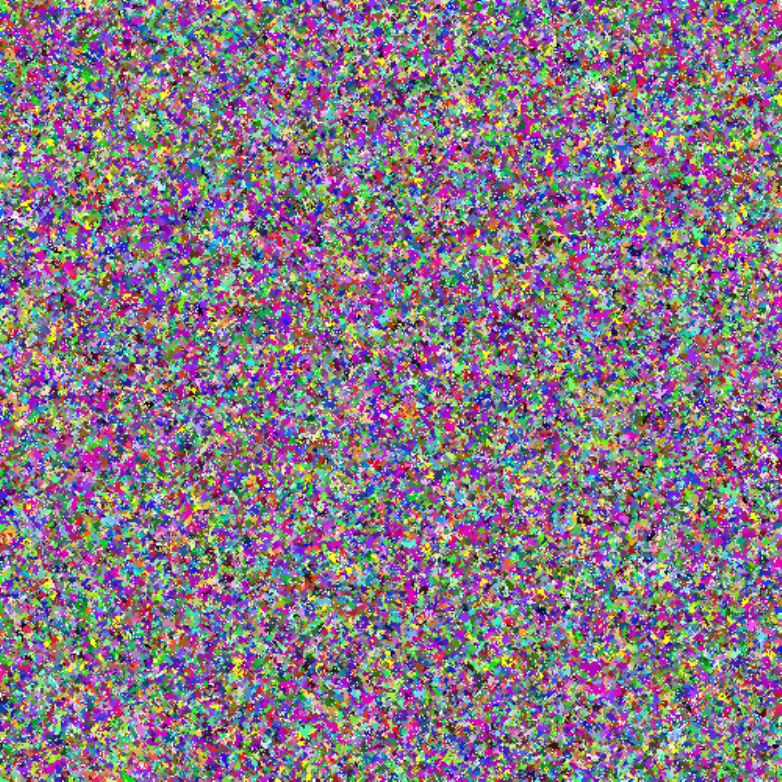}
            
            \label{fig:genomes500}
        \end{subfigure}%
        \begin{subfigure}[b]{0.19\textwidth}
            \includegraphics[width=0.93\linewidth]{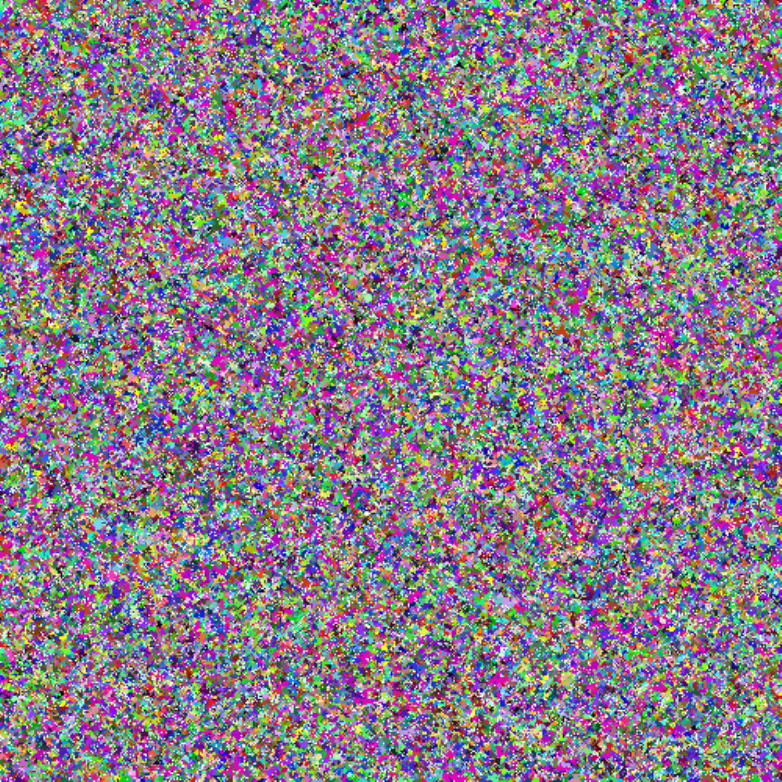}
            
            \label{fig:genomes750}
        \end{subfigure}%
        \begin{subfigure}[b]{0.19\textwidth}
            \includegraphics[width=0.93\linewidth]{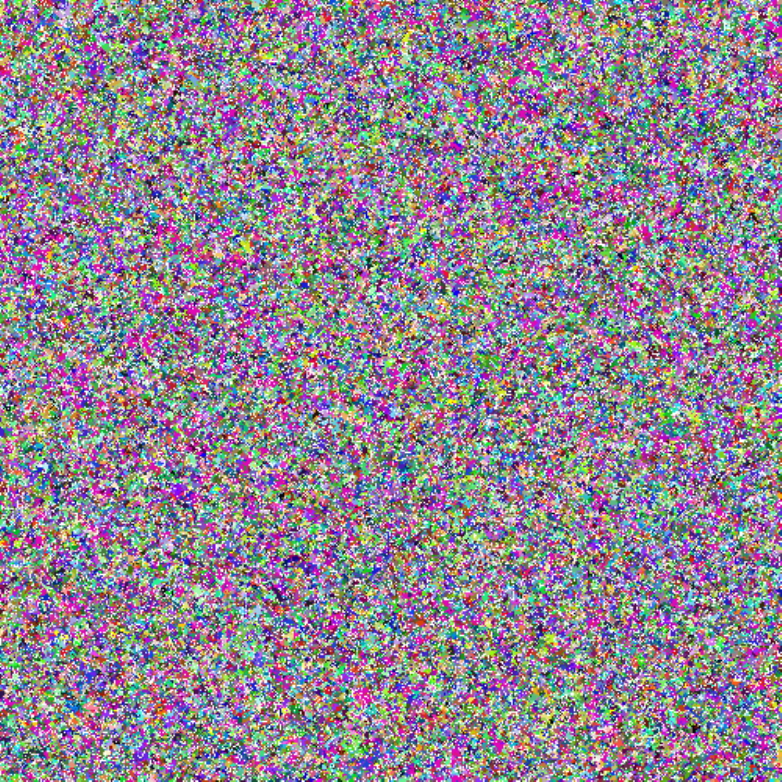}
            
            \label{fig:genomes1000}
        \end{subfigure}
    
        \medskip
    
        \begin{subfigure}[b]{0.19\textwidth}
            \includegraphics[width=0.93\linewidth]{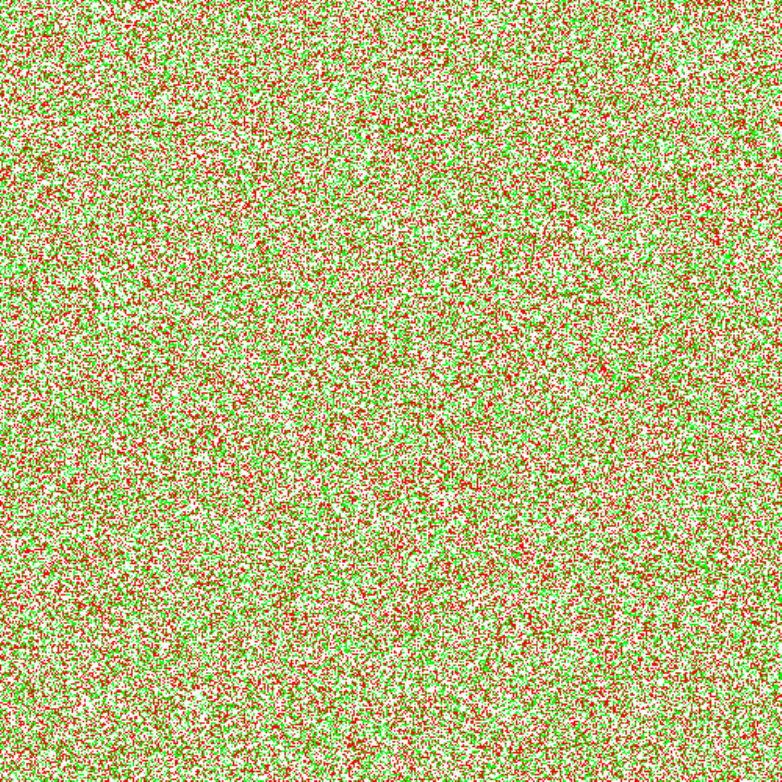}
            
            \label{fig:gridstates1}
        \end{subfigure}%
        \begin{subfigure}[b]{0.19\textwidth}
            \includegraphics[width=0.93\linewidth]{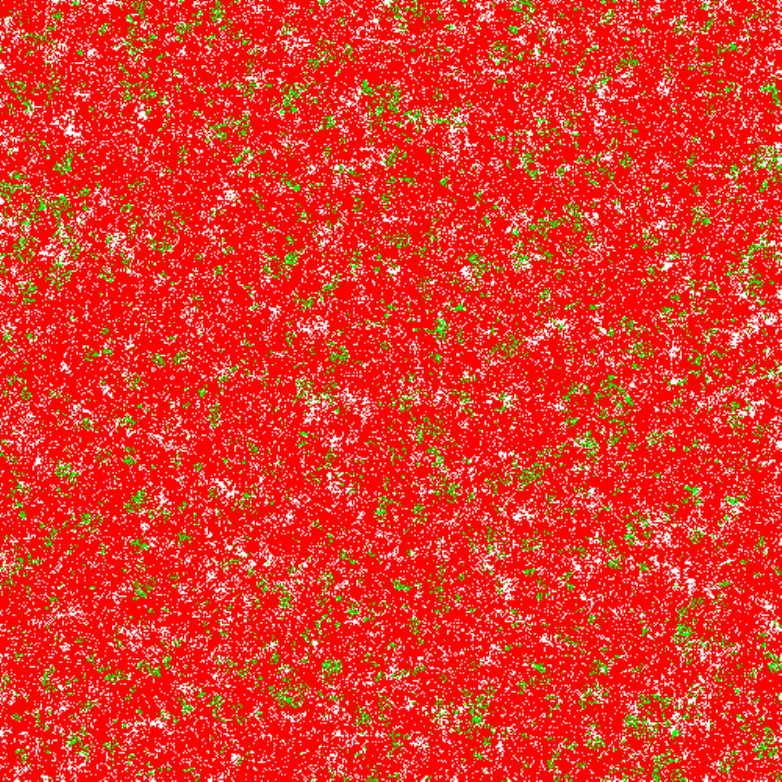}
            
            \label{fig:gridstates250}
        \end{subfigure}%
        \begin{subfigure}[b]{0.19\textwidth}
            \includegraphics[width=0.93\linewidth]{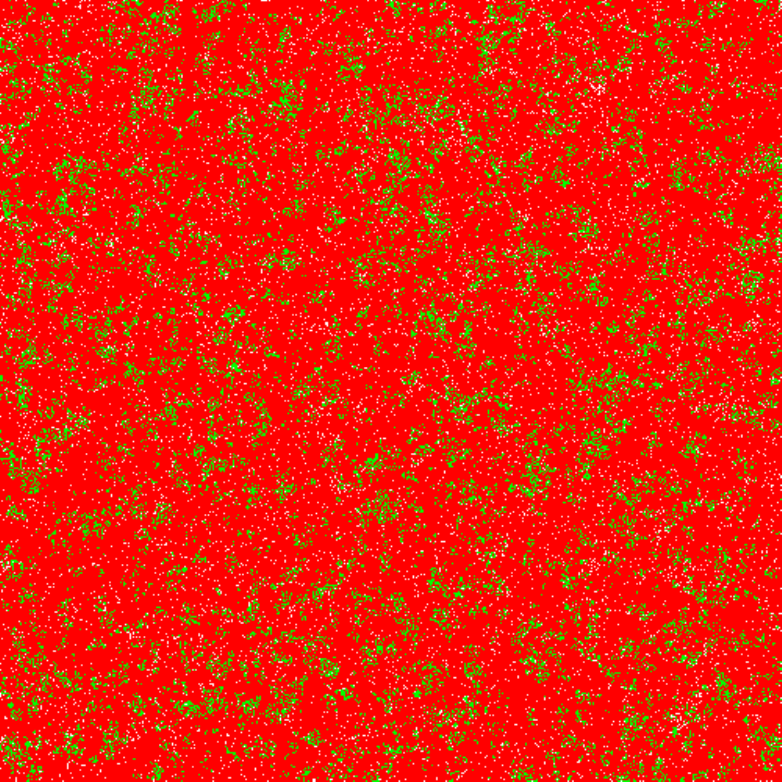}
            
            \label{fig:gridstates500}
        \end{subfigure}%
        \begin{subfigure}[b]{0.19\textwidth}
            \includegraphics[width=0.93\linewidth]{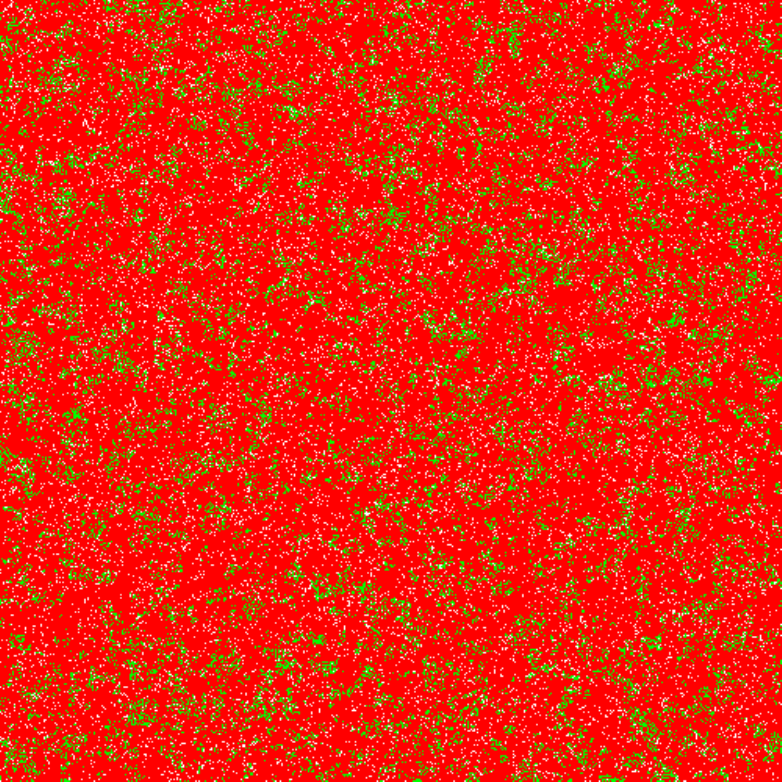}
            
            \label{fig:gridstates750}
        \end{subfigure}%
        \begin{subfigure}[b]{0.19\textwidth}
            \includegraphics[width=0.93\linewidth]{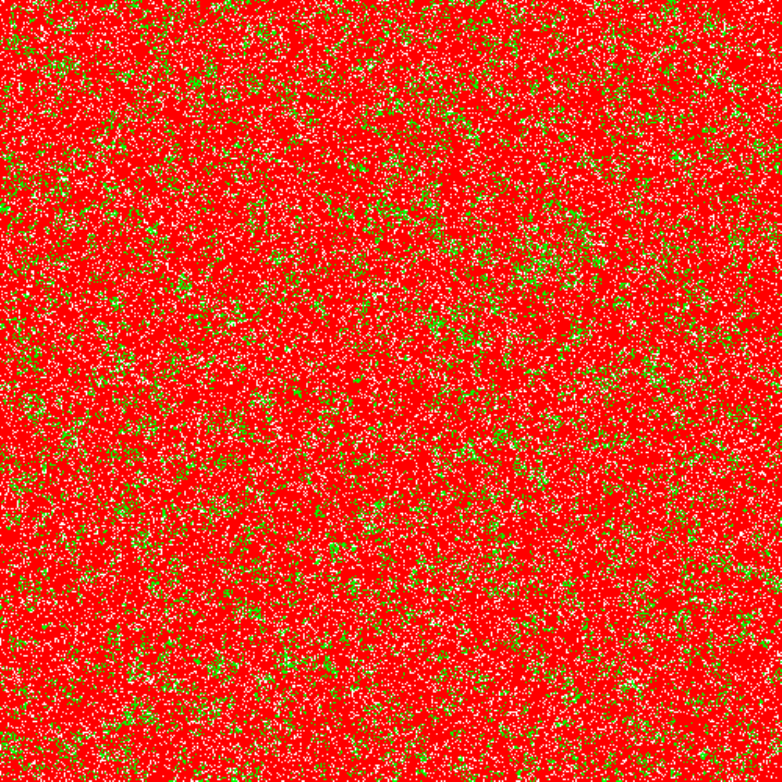}
            
            \label{fig:gridstates1000}
        \end{subfigure}
    
        \caption{Example execution of 500 $\times$ 500 grid, $a_{max}$ = 50, $a_{dec}$ = 70. First Row: \textit{Cell States}, Second Row: Genomes, and Third Row: \textit{Grid States} - Columns: Snapshots at generation 1, 250, 500, 750, 1,000.}
        \label{fig:5005001k_5070}
    \end{figure*}

The tiles in \ref{fig:505010k_1015}, \ref{fig:505010k_5070}, \ref{fig:5005001k_1015}, and \ref{fig:5005001k_5070} qualitatively illustrate the behaviour of the CA over the generations in a single run. The first row of the plots shows the almost constant amount of alive, decaying and dead cells in later generations, while the actual positions of those states change over time. In the second row, the genome of the cells is color-coded. Of course, the combinational possibilities make it impossible to have a humanly recognizable unique color for every genome, but it can be seen how there are small populations of similar genomes spread over the grid with no single population dominating it and with dead cells that allow for (re)growth in between. Especially for the longer lifespan, the convergence of cell with state $0$ and $1$ and the increasing amount of cell with state $1$ can be observed in the lower row.

\section{Conclusion and Future Work}

We have demonstrated an heterogeneous life-like CA, where long-term phenotypic dynamics are possible thanks to an evolutionary inner loop. Such long-term behaviors are an important ingredient for open-endedness, which is typically lacking in other homogeneous life-like CA such as the Game of Life. Additionally, in this work we incorporate the concept of cell ageing to ensure that cell aliveness and actual cell computation are kept conceptually separated. We plan to conduct further studies where cell ageing can be perturbed by an environment or a task to be solved, and the substrate should show further adaptation.


\begin{credits}
\subsubsection{\ackname} This work was partially financed by the Research Council of Norway's DeepCA project, grant agreement 286558.
\end{credits}

%
%
%
\bibliographystyle{splncs04}
\bibliography{example}
\newpage
\section{Appendix - Figures}

\end{document}